%% file: Paper.tex
\documentclass[cmp,numbook]{svjour} 
\usepackage{graphics}
\usepackage{times}
\usepackage{amsmath,amssymb}
\usepackage[dvips]{graphicx, color}
\usepackage{verbatim}
\usepackage{bbm}
\usepackage{longtable}
\usepackage{amscd}
\usepackage{enumerate}
\usepackage[all]{xy}
\usepackage{stmaryrd}

\DeclareMathOperator{\tr}{tr}

\DeclareMathOperator{\sgn}{sgn}

\DeclareMathOperator{\Sym}{Sym}
\DeclareMathOperator{\Sph}{\mathbb{S}}

\DeclareMathOperator{\End}{End}
\DeclareMathOperator{\Lie}{{\mathfrak g}}

\newcommand{\XX}{\sigma} 
\newcommand{\YY}{\tau} 

\newcommand{\dd}{\mathrm{d}}

\newcommand{\vertical}{\perp} 
\newcommand{\vacuum}{\text{vac}}

\newcommand{\R}{\mathbb{R}}
\newcommand{\C}{\mathbb{C}} 
\newcommand{\Z}{\mathbb{Z}}

\newcommand{\p}{\partial}

\newcommand{\met}{g}
\newcommand{\cmet}{{\mathfrak H}}
\newcommand{\cmetrep}{{\mathfrak h}}

\renewcommand{\sb}[2]{\llbracket #1,#2\rrbracket}

\newcommand{\conn}{\mathcal{P}}

\newcommand{\dmd}{\Diamond} 
\newcommand{\fdmd}{\blacklozenge}

\newcommand{\antisym}{\mathbf{A}}

\spnewtheorem{convention}{Convention}[section]{\it}{\rm}

\makeatletter

\newcommand{\Rmnum}[1]{\expandafter\@slowromancap\romannumeral #1@}
\makeatother

\newcommand{\slaz}[1]{#1\hskip-4.5pt/\hskip1pt}
\newcommand{\slaa}[1]{#1\hskip-5.5pt/\hskip1pt}
\newcommand{\slab}[1]{#1\hskip-7pt/\hskip1pt}

\newcommand{\diffeo}{\rho}

\begin{document}

\rule{0pt}{0pt}
\vskip 40mm 
{\bf \Large
\noindent A formalism for analyzing vacuum spacetimes
}
\vskip 4mm
\noindent {\bf Michael Reiterer, Eugene Trubowitz}
\vskip 2mm
\noindent {Department of Mathematics, ETH Zurich, Switzerland}
\vskip 4mm
\noindent {\bf Abstract:} The Einstein vacuum equations in the formulation developed by Newman, Penrose \cite{NP} and Friedrich \cite{Fr} are expressed in terms of a Lie superbracket. Differential identities are derived from the super Jacobi identity.
This perspective clarifies the covariance properties of the equations. The equations are intended as a tool for the analytic  study of vacuum spacetimes.

\input{SectionDD.tex}

\input{SectionDiamonds}
\input{SectionLorentzian.tex}

\input{SectionReform.tex}
\input{SectionIndices.tex}
%

%
%

\end{document}

%% file: SectionDD.tex
\section{Introduction}
In this paper, we discuss a formalism that is suited to the analysis of solutions to the Einstein vacuum equations. In this formalism, the vacuum equations
\begin{itemize}
\item become a quasilinear, first order system of partial differential equations, that
\item are quadratically nonlinear, and
\item through gauge-fixing, can be brought into symmetric hyperbolic form.
\end{itemize}
Newman and Penrose \cite{NP} introduced the basic unknown fields of this formalism (frame, connection, Weyl curvature) and the corresponding Einstein vacuum equations. Their equations are not independent, but satisfy general differential identities, that were derived by Friedrich \cite{Fr}.

Friedrich \cite{Fr} showed, by choosing an appropriate gauge, that the vacuum equations contain a symmetric hyperbolic subsystem that determines the evolution of all unknown fields. To show that the remaining equations, called \emph{constraints}, are also fulfilled, he used the general differential identities. 

In this paper, the vacuum equations as formulated by Newman and Penrose are expressed in terms of a Lie superbracket, see \eqref{eq:ljflksdjlsdjs} and \eqref{eq:hkjhkhkfds}. The general differential identities, see \eqref{eq:mn2}, are derived from the associated super Jacobi identity. We take special care to exhibit the covariance properties of the equations.


We used a forerunner of the present formalism to analyze strongly focused gravitational waves, see Appendix B of \cite{RT}.
The point of the refined presentation of this paper is the derivation of the equations in Section \ref{sec:instrman} from an invariant point of view. They are intended to be used as a tool in the analysis of other problems in classical general relativity.

\emph{Important remark:} We expect that there is a close relationship between the notion of a Cartan connection, see \cite{Sh}, and the formalism of this paper, which is not made here. This relationship ought to be clarified.
However, we have not pursued this relationship, since the equations of Section \ref{sec:instrman} can be derived without it.

\section{A Lie Superalgebra Identity}\label{sec:abstractidentity}
We recall the definition of a real Lie superalgebra:
\begin{definition} \label{def:spp}
A (non-associative) $\Z_2$-graded real algebra $\big(L=L_0\oplus L_1,\sb{\,\cdot\,}{\,\cdot\,}\big)$, with even parts $L_0$ and odd parts $L_1$, satisfying for all $x_1 \in L_{k_1}$, $x_2 \in L_{k_2}$, $x_3 \in L_{k_3}$
\begin{enumerate}[(a)]
\item\label{enum0071} $\sb{x_1}{x_2} \in L_{\ell}$ with $\ell = k_1+k_2 \pmod{2}$
\item\label{enum0072} $\sb{x_1}{x_2} = (-1)^{1+k_1k_2} \sb{x_2}{x_1}$
\item\label{enum0073} $(-1)^{k_1k_3} \sb{x_1}{\sb{x_2}{x_3}} + (-1)^{k_2k_1}\sb{x_2}{\sb{x_3}{x_1}} + (-1)^{k_3k_2}\sb{x_3}{\sb{x_1}{x_2}} = 0$
\end{enumerate}
is called a \emph{real Lie superalgebra}. In this context, $\sb{\,\cdot\,}{\,\cdot\,}$ is the Lie superbracket, (\ref{enum0072}) is super skew-symmetry, and (\ref{enum0073}) is the super Jacobi identity.
\end{definition}
Let $\big(L=L_0\oplus L_1,\sb{\,\cdot\,}{\,\cdot\,}\big)$ be a Lie superalgebra as above. Set $A_0 = L_0 \times L_1$ and $A_1 = L_1 \times L_0$, that is
$A_{\ell} = L_{\ell}\times L_{\ell+1}$ for all $\ell \in \Z_2$.
\begin{definition} For $\ell \in \Z_2$, let
\begin{alignat*}{2}
\mathcal{D}^{(\ell)}: A_1 \times A_{\ell} & \to A_{\ell+1} &\qquad (x,y) & \mapsto \mathcal{D}^{(\ell)}_x y
\end{alignat*}
where
\begin{equation} \label{eq:ddd3}
\mathcal{D}^{(\ell)}_xy = z = (z_1,z_2) \in A_{\ell+1}
\qquad \text{with} \qquad \left \{
\begin{aligned}
z_1 & = y_2 - \epsilon_{\ell} \sb{x_1}{y_1}\\
z_2 & = \epsilon_{\ell} \sb{x_1}{y_2} + \epsilon_{\ell}\sb{y_1}{x_2}
\end{aligned}\right.
\end{equation}
for all $x = (x_1,x_2) \in A_1$ and all $y = (y_1,y_2) \in A_{\ell}$. Here, $\epsilon_0 = 1$ and $\epsilon_1 = \tfrac{1}{2}$.
\end{definition}
Equation \eqref{eq:ddd3} is consistent, because $x_1 \in L_1$, $x_2 \in L_0$, $y_1 \in L_{\ell}$, $y_2\in L_{\ell+1}$ imply $z_1 \in L_{\ell+1}$, $z_2\in L_{\ell}$, as required.
\begin{convention}
From now on, we will drop the superscripts $(0)$, $(1)$ on the operator $\mathcal{D}$, with the understanding that "the arguments determine the superscript".
\end{convention}
\begin{proposition} \label{prop:hhhgd} $\mathcal{D}_x \mathcal{D}_x x = 0$ for all $x\in A_1$
\end{proposition}
\begin{proof} 
Let $y = \mathcal{D}_xx$ and $z = \mathcal{D}_xy$. We have to show that $z=0$. We have
\begin{subequations}
\begin{align}
\label{eq:ggggggr1} y_1 & = x_2 - \tfrac{1}{2} \sb{x_1}{x_1} \\
\label{eq:ggggggr2} y_2 & = \sb{x_1}{x_2}\\
\intertext{and therefore}
\label{eq:ggggggr3} z_1 & =  y_2 - \sb{x_1}{y_1}  = \tfrac{1}{2} \sb{x_1}{\sb{x_1}{x_1}} \\
\label{eq:ggggggr4} z_2 & = \sb{x_1}{y_2} + \sb{y_1}{x_2}
= \sb{x_1}{\sb{x_1}{x_2}}  - \tfrac{1}{2}\sb{\sb{x_1}{x_1}}{x_2} + \sb{x_2}{x_2}
\end{align}
\end{subequations}
Recalling that $x_1 \in L_1$ and $x_2 \in L_0$, the super skew symmetry (\ref{enum0072}) and the super Jacobi identity (\ref{enum0073}) in Definition \ref{def:spp} imply $\sb{x_1}{\sb{x_1}{x_1}} = 0$, $\sb{x_2}{x_2} = 0$ and $\sb{\sb{x_1}{x_1}}{x_2} = 2\sb{x_1}{\sb{x_1}{x_2}}$. For example,
\begin{align*}
0 & = \sb{x_1}{\sb{x_1}{x_2}} + \sb{x_2}{\sb{x_1}{x_1}} - \sb{x_1}{\sb{x_2}{x_1}}\\
 & = \sb{x_1}{\sb{x_1}{x_2}} - \sb{\sb{x_1}{x_1}}{x_2} + \sb{x_1}{\sb{x_1}{x_2}}
\end{align*}
Therefore, $z = 0$. \qed
\end{proof}
\begin{remark}
In Section \ref{sec:llllllllla21} the abstract equation $\mathcal{D}_x x = 0$ for the unknown "field" $x \in A_1$ will be interpreted as "Einstein vacuum equations". There are too many equations. The system is apparently overdetermined. The remedy is the identity of Proposition \ref{prop:hhhgd}, that holds for all $x\in A_1$.
\end{remark}

%% file: SectionDiamonds.tex
\section{Diamonds}\label{sec:diamonds}
\begin{convention} In this paper, all manifolds are real, smooth and finite dimensional.
For any fiber bundle $\pi: E \to B$, the fiber over $p\in B$ is denoted by $E_p = \pi^{-1}(\{p\})$. For any section $X \in \Gamma(E)$ the map $X: B \to E$ is given by $p \mapsto X_p \in E_p$. For any vector bundle $\pi: E \to B$ we denote by $E^{\ast}$, $\Sym^2 E$, $\Sph E$, the dual bundle, the subbundle of symmetric elements of $E\otimes E$, and the sphere bundle associated with $E$. That is, for $p\in B$, we have $(\Sph E)_p = (E_p \setminus \{0\})/\R_+$.
Finally, $\End(E) = E^{\ast}\otimes E$ is the endomorphism bundle associated with $E$.
\end{convention}
\begin{convention} \label{conv:uzt} For a bundle $\pi: E \to B$ we denote by $\mathcal{T}(E)$ the algebraic direct sum of all tensor products of $E$ and $E^{\ast}$.
\end{convention}
For the {\bf rest of this paper}, fix
\begin{itemize}
\item a 4-dimensional manifold $M$,
\item a real vector bundle $\pi_V: V \to M$ with 4 dimensional fibers,
\item a section $\cmet \in \Gamma(\Sph \Sym^2 V^{\ast})$ with signature $(-,+,+,+)$.
\end{itemize}
In other words, $\cmet$ defines a \emph{conformal} Lorentzian inner product on each fiber of $V$.
\begin{definition}\label{def:mnnnbvc} For every integer $k\geq 0$, let $\conn^k$ be the set of all maps $\dmd$,
\begin{equation}\label{eq:jahgs}
\dmd:\; \Gamma(\mathcal{T}(V)) \to \Gamma\big({\wedge^k V^{\ast}} \otimes \mathcal{T}(V)\big)
\end{equation}
so that for all $u,v \in \Gamma(\mathcal{T}(V))$, all representatives $\cmetrep \in \Gamma(\Sym^2 V^{\ast})$ of the conformal Lorentzian inner product $\cmet \in \Gamma(\Sph \Sym^2 V^{\ast})$, and all $Y\in \Gamma(V^{\otimes k})$, we require, with Convention \ref{conv:jjjjjutg} below:
\begin{enumerate}[(a)]
\item\label{tre1} $\dmd$ is linear over $\R$,
\item\label{trexy} $\dmd_Y$ maps $C^{\infty}(M) \to C^{\infty}(M)$ and $\Gamma(V)\to \Gamma(V)$ and $\Gamma(V^{\ast})\to\Gamma(V^{\ast})$,
\item\label{tre2} $\dmd_Y(u \otimes v) = (\dmd_Y u)\otimes v + u \otimes (\dmd_Y v)$,
\item\label{trenew} $\dmd I = 0$ if $I \in \Gamma(\End(V))$ is the identity on the fibers of $V$,
\item\label{tre3} $\dmd \cmetrep = \mu \otimes \cmetrep$ for some $\mu \in \Gamma({\wedge^k V^{\ast}})$.
\end{enumerate}
The vertical subspace $\conn^k_{\vertical} \subset \conn^k$ is the set of all $\dmd\in \conn^k$ such that $\dmd f = 0$ for all $f \in C^{\infty}(M)$.
\end{definition}
\begin{convention}\label{conv:jjjjjutg}
For each  $Y \in \Gamma(V^{\otimes k})$ and $u \in \Gamma(\mathcal{T}(V))$ set
$$\dmd_Yu = i_Y\big(\dmd u\big) \;\in \; \Gamma(\mathcal{T}(V))$$
Here $i_Y$ is interior multiplication by $Y$ acting on the first $k$ factors of $\dmd u$.
\end{convention}
\begin{remark} Observe that $\dmd_Y$ acts on the ring $C^{\infty}(M)$ as a derivation, by (\ref{tre2}).
\end{remark}
\begin{remark} \label{rem_klioio}
Every element of $\conn^k$ can be written as a finite sum of "pure" elements $\theta \otimes \dmd$, where $\theta \in \Gamma({\wedge^k} V^{\ast})$ and $\dmd \in \conn^0$. The Leibniz rule (c) for $\theta \otimes \dmd$ reads
$$\big(\theta \otimes \dmd)(u\otimes v) = \theta \otimes (\dmd u) \otimes v + \theta \otimes u \otimes (\dmd v)$$
\end{remark}
\begin{remark} \label{rem:jhkjhks}
Let $\mathcal{I}$ be an index set, $|\mathcal{I}|=4$. Let $F_{(a)}$, $a \in \mathcal{I}$, be local sections of $V$ that are a frame for fibers of $V$. Let $\lambda^{(a)}, a \in \mathcal{I}$, be the dual frame. For every $\dmd \in \conn^k$
and $Y \in \Gamma(V^{\otimes k})$, property (\ref{trenew}) in Definition \ref{def:mnnnbvc} implies
$$0 = \dmd_YI = \dmd_Y\big(\lambda^{(a)}\otimes F_{(a)}\big)
= \big(\dmd_Y \lambda^{(a)}\big)\otimes F_{(a)} + \lambda^{(a)}\otimes \big(\dmd_Y F_{(a)}\big)$$
Consequently, $(\dmd_Y \lambda^{(a)})(F_{(b)}) = - \lambda^{(a)}\big(\dmd_Y F_{(b)}\big)$
for all $a,b\in \mathcal{I}$. Now, the Leibniz rule (\ref{tre2}) implies
\begin{equation}\label{eq:jkjkjlfhkhf}
(\dmd_Y \xi)(Z) = \dmd_Y(\xi(Z)) - \xi(\dmd_YZ)
\end{equation}
for all $\xi \in \Gamma(V^{\ast})$ and $Z \in \Gamma(V)$.
\end{remark}
\begin{definition}
Let $m, k_1,\ldots,k_{\ell} \geq 0$ be integers, and $k=k_1 + \ldots + k_{\ell}$. The multi $(k_1,\ldots,k_{\ell})$ wedge product operator shifted by $m$ is the linear map
\begin{align*}
\wedge^{(m)}_{k_1,\ldots,k_{\ell}}:\;  \Gamma\Big(({\wedge^m} V^{\ast})\otimes ({\wedge^{k_1}V^{\ast}})
\otimes \cdots & \otimes ({\wedge^{k_{\ell}}V^{\ast}}) \otimes \mathcal{T}(V) \Big)\\
& \to \Gamma\Big(({\wedge^m} V^{\ast})\otimes  (\wedge^kV^{\ast}) \otimes \mathcal{T}(V)\Big)
\end{align*}
determined by $\xi \otimes \nu_1 \otimes \cdots \otimes \nu_{\ell} \otimes u  \mapsto \xi \otimes \big(\nu_1 \wedge \cdots \wedge \nu_{\ell}\big) \otimes u$. Set $\wedge_{k_1,\ldots,k_{\ell}} = \wedge^{(0)}_{k_1,\ldots,k_{\ell}}$.
\end{definition}
\begin{remark} \label{rem:khdhfskhfdsk} We have
\begin{align*}
{\wedge_{k_1,k_2+k_3}} \wedge_{k_2,k_3}^{(k_1)} & = {\wedge_{k_1,k_2,k_3}}\\
\dmd \wedge_{k_2,k_3} & =\wedge_{k_2,k_3}^{(k_1)} \dmd
\end{align*}
for any $\dmd \in \conn^{k_1}$.
\end{remark}
\begin{proposition}\label{prop:ksaw2}
For all $\dmd \in \conn^k$, $
\dmd\hskip-6pt/\hskip1pt \in \conn^{\ell}$, set
\begin{equation} \label{eq:kiijijid}
\sb{\dmd}{\dmd\hskip-5.5pt/\hskip1pt} = 
 \wedge_{k,\ell}\dmd \dmd\hskip-5.5pt/\hskip1pt - (-1)^{k\ell} \wedge_{\ell,k} 
\dmd\hskip-5.5pt/\hskip1pt \dmd
\end{equation}
Then $\sb{\dmd}{\dmd\hskip-5.5pt/\hskip1pt} \in \conn^{k+\ell}$ and moreover, $\big(\conn_0\oplus \conn_1,\sb{\,\cdot\,}{\,\cdot\,}\big)$ is a Lie superalgebra, with $\conn_0 = \bigoplus_{k\geq 0\,\text{even}} \conn^k$
and $\conn_1 = \bigoplus_{k\geq 0\,\text{odd}} \conn^k$.
\end{proposition}
\begin{proof}
To see that $\sb{\dmd}{\slaa{\dmd}} \in \conn^{k+\ell}$, consider first the special case when $k=\ell = 0$.
In this case $\sb{\dmd}{\slaa{\dmd}} = \dmd\slaa{\dmd} - \slaa{\dmd}\dmd$. 
Properties (a), (b), (d) in Definition \ref{def:mnnnbvc} hold. The Leibniz rule (c) holds:
\begin{align*}
\sb{\dmd}{\slaa{\dmd}} (u\otimes v) & = \dmd \slaa{\dmd}(u\otimes v) - \slaa{\dmd}\dmd (u\otimes v)\\
& = \dmd \big((\slaa{\dmd} u)\otimes v\big)
+ \dmd \big(u\otimes (\slaa{\dmd} v)\big) - \slaa{\dmd}\big((\dmd u) \otimes v\big) - \slaa{\dmd}\big(u\otimes (\dmd v)\big)\\
& =
(\dmd \slaa{\dmd} u)\otimes v
+ (\slaa{\dmd} u)\otimes (\dmd v)
+(\dmd u)\otimes (\slaa{\dmd} v)
+ u\otimes (\dmd \slaa{\dmd} v)\\
& \qquad  - (\slaa{\dmd}\dmd u) \otimes v
 - (\dmd u) \otimes (\slaa{\dmd}v)
 -( \slaa{\dmd} u) \otimes (\dmd v)
 - u\otimes ( \slaa{\dmd}\dmd v)\\
& = \big(\sb{\dmd}{\slaa{\dmd}}u\big)\otimes v + u \otimes \big(\sb{\dmd}{\slaa{\dmd}} v\big)
\end{align*}
For property (e), note that there are $\mu,\slaa{\mu} \in \C^{\infty}(M)$ such that $\dmd \cmetrep = \mu \cmetrep$
and $\slaa{\dmd} \cmetrep = \slaa{\mu} \cmetrep$.  
$$\sb{\dmd}{\slaa{\dmd}} \cmetrep = \dmd (\slaa{\mu} \cmetrep) - \slaa{\dmd}( \mu \cmetrep)
= (\dmd \slaa{\mu}) \cmetrep + \slaa{\mu} \mu \cmetrep - (\slaa{\dmd}\mu)\cmetrep - \mu \slaa{\mu}\cmetrep = 
\big(\dmd \slaa{\mu} - \slaa{\dmd}\mu\big)\cmetrep$$
Therefore, (e) holds. For general $k,\ell$, (a), (b) and (d) still hold.  For the Leibniz rule (c), observe that both sides of \eqref{eq:kiijijid} are bilinear over $\R$ in $\dmd$ and 
$\slaa{\dmd}$. It therefore suffices to consider the case when $\dmd = \theta\otimes \dmd_0$ and $\slaa{\dmd}
= \slaa{\theta} \otimes \slaa{\dmd}_0$, where $\dmd_0,\slaa{\dmd}_0 \in \mathcal{P}^0$ and $\theta \in \Gamma(\wedge^k V^{\ast})$ and $\slaa{\theta} \in \Gamma({\wedge^{\ell}V^{\ast}})$.
In this case,
\begin{align}
\notag \sb{\dmd}{\slaa{\dmd}} & = \wedge_{k,\ell}\; \theta \otimes \dmd_0 \big( \slaa{\theta}\otimes \slaa{\dmd}_0\big)
- (-1)^{k\ell} \wedge_{\ell,k}\,\slaa{\theta} \otimes \slaa{\dmd}_0 \big(\theta \otimes \dmd_0\big)\\
\label{eq:klklkld} & = 
(\theta \wedge \slaa{\theta}) \otimes  \sb{\dmd_0}{\slaa{\dmd}_0}
+ \big(\theta \wedge (\dmd_0 \slaa{\theta})\big) \otimes \slaa{\dmd}_0
- \big((\slaa{\dmd}_0\theta)\wedge \slaa{\theta}\big) \otimes \dmd_0
\end{align}
Each term separately satisfies the Leibniz rule (the first one by the special case $k=\ell = 0$), and (c) holds. 
Property (e) also follows from \eqref{eq:klklkld}.\\
To see that $\sb{\,\cdot\,}{\,\cdot\,}: \conn^k \times \conn^{\ell} \to \conn^{k+\ell}$ is a Lie superbracket, observe that
\begin{align*}
\sb{\slaa{\dmd}}{\dmd} & = \wedge_{\ell,k} \slaa{\dmd}\dmd - (-1)^{k\ell} \wedge_{k,\ell}\dmd\slaa{\dmd}\\
& = (-1)^{1+k\ell} \big(\wedge_{k,\ell} \dmd\slaa{\dmd} - (-1)^{k\ell} \wedge_{\ell,k}\slaa{\dmd}\dmd\big) \\
& = (-1)^{1+k\ell} \sb{\dmd}{\slaa{\dmd}}
\end{align*}
Let $\dmd_1 \in \conn^{k_1}$, $\dmd_2 \in \conn^{k_2}$, $\dmd_3 \in \conn^{k_3}$. Then
\begin{align*}
\sb{\dmd_1}{\sb{\dmd_2}{\dmd_3}}
& = {\wedge_{k_1,k_2+k_3}} \dmd_1 \wedge_{k_2,k_3} \dmd_2\dmd_3
- (-1)^{k_2k_3} {\wedge_{k_1,k_2+k_3}} \dmd_1 \wedge_{k_3,k_2} \dmd_3\dmd_2\\
& \qquad - (-1)^{k_1(k_2+k_3)} {\wedge_{k_2+k_3,k_1}}
{\wedge_{k_2,k_3}} \dmd_2\dmd_3 \dmd_1\\
& \qquad + (-1)^{k_1(k_2+k_3)+k_2k_3} {\wedge_{k_2+k_3,k_1}}
 \wedge_{k_3,k_2} \dmd_3\dmd_2 \dmd_1
\end{align*}
By Remark \ref{rem:khdhfskhfdsk},
\begin{align*}
& (-1)^{k_1k_3} \sb{\dmd_1}{\sb{\dmd_2}{\dmd_3}}\\
& = (-1)^{k_1k_3} {\wedge_{k_1,k_2,k_3}} \dmd_1 \dmd_2\dmd_3
- (-1)^{k_1k_2} {\wedge_{k_2,k_3,k_1}} \dmd_2\dmd_3 \dmd_1\\
& \qquad - (-1)^{k_3(k_1+k_2) } {\wedge_{k_1,k_3,k_2}} \dmd_1 \dmd_3\dmd_2
 + (-1)^{k_2(k_1+k_3)} {\wedge_{k_3,k_2,k_1}}  \dmd_3\dmd_2 \dmd_1
\end{align*}
Adding,
$$(-1)^{k_1k_3} \sb{\dmd_1}{\sb{\dmd_2}{\dmd_3}}
+ (-1)^{k_2k_1} \sb{\dmd_2}{\sb{\dmd_3}{\dmd_1}}
+ (-1)^{k_3k_2} \sb{\dmd_3}{\sb{\dmd_1}{\dmd_2}} = 0$$
 \qed
\end{proof}
\begin{convention}\label{conv:mkxs} The symbol $\mathcal{J}$ denotes a finite index set. The set $\mathcal{J}$ and its length $|\mathcal{J}|$ \emph{may change from occurrence to occurrence}. Boldface small Latin indices $\mathbf{a}, \mathbf{b},\ldots $ take values in $\mathcal{J}$. Boldface Capital Latin indices are multiindices, that is, elements of $\mathcal{J}^k$ for some $k\geq 0$. The length of a multiindex $\mathbf{A} = (\mathbf{a}_1,\ldots,\mathbf{a}_k)$ will be denoted $|\mathbf{A}| = k$. We write $X_{\mathbf{A}} = X_{\mathbf{a}_1}\otimes \cdots \otimes X_{\mathbf{a}_k}$, for various types of objects $X$.
\end{convention}
\begin{definition}\label{def:hzttr}
Let $\mathcal{J}$ be an index set and let $\mathbf{A}$, $\mathbf{B}_1$, \ldots, $\mathbf{B}_{\ell}$ be $\mathcal{J}$-multiindices such that $|\mathbf{A}| = |\mathbf{B}_1| + \ldots + |\mathbf{B}_{\ell}| = k$.
Let $\mathbf{A} = (\mathbf{a}_1,\ldots,\mathbf{a}_k)$ and let $\mathbf{B}_1||\cdots || \mathbf{B}_{\ell}
= (\mathbf{b}_1,\ldots,\mathbf{b}_k)$ be the concatenation of $\mathbf{B}_1$ through $\mathbf{B}_{\ell}$. Set
\begin{equation}\label{eq:jjjhztgfred}
{\antisym_{\mathbf{A}}}^{\mathbf{B}_1\cdots \mathbf{B}_{\ell}} = \tfrac{1}{|\mathbf{B}_1|!\,\cdots\, |\mathbf{B}_{\ell}|!}  \sum_{\pi \in S_k} \sgn(\pi) {\delta_{\mathbf{a}_{\pi(1)}}}^{\mathbf{b}_1}
\cdots {\delta_{\mathbf{a}_{\pi(k)}}}^{\mathbf{b}_k}
\end{equation}
The index set $\mathcal{J}$ is implicit in 
\eqref{eq:jjjhztgfred} and will be specified every time it is used.
\end{definition}
\begin{remark} ${\antisym_{\mathbf{A}}}^{\mathbf{BC}}{\antisym_{\mathbf{B}}}^{\mathbf{DE}} = {\antisym_{\mathbf{A}}}^{\mathbf{DEC}}$ where $|\mathbf{A}| = |\mathbf{B}| + |\mathbf{C}|
= |\mathbf{D}| + |\mathbf{E}| + |\mathbf{C}|$.
\end{remark}
\begin{remark} \label{rem:kiiiiujh} Let $\dmd \in \conn^k$, $Y \in \Gamma(V^{\otimes k})$ and $z \in \Gamma(\mathcal{T}(V))$.
Then
\begin{equation}\label{eq:loiy2}
[z\,\otimes\,,\dmd_Y] =
- (\dmd_Y z)\,\otimes
\end{equation}
as operators acting on $\Gamma(\mathcal{T}(V))$, and $[\,\cdot\,,\,\cdot\,]$ is the commutator of operators.
\end{remark}
\begin{remark} \label{rem:jkjkjiu}
Equation \eqref{eq:kiijijid} is equivalent to
\begin{subequations}
\begin{align}
\notag & i_{Y_{\mathbf{A}}} \sb{\dmd}{\dmd\hskip-5.5pt/\hskip1pt}  \\
\label{rfgtr1} & =  {\antisym_{\mathbf{A}}}^{\mathbf{BC}} 
\Big(i_{Y_{\mathbf{B}}\otimes Y_{\mathbf{C}}} \dmd \dmd\hskip-5.5pt/\hskip1pt
- i_{Y_{\mathbf{C}} \otimes Y_{\mathbf{B}}} \dmd\hskip-5.5pt/\hskip1pt \dmd\Big)\\
\label{rfgtr2}   & =  {\antisym_{\mathbf{A}}}^{\mathbf{BC}} 
\big(\dmd_{Y_{\mathbf{B}}}\dmd\hskip-5.5pt/\hskip1pt_{Y_{\mathbf{C}}}
-\dmd\hskip-5.5pt/\hskip1pt_{Y_{\mathbf{C}}}\dmd_{Y_{\mathbf{B}}}\big)
-  {\antisym_{\mathbf{A}}}^{\mathbf{BcE}} \dmd\hskip-5.5pt/\hskip1pt_{(\dmd_{Y_{\mathbf{B}}}Y_{\mathbf{c}}) \otimes Y_{\mathbf{E}}} 
+ {\antisym_{\mathbf{A}}}^{\mathbf{bDC}} \dmd_{(\dmd\hskip-4.5pt/\hskip1pt_{Y_{\mathbf{C}}} Y_{\mathbf{b}}) \otimes Y_{\mathbf{D}}}
\end{align}
\end{subequations}
Here $Y_1,\ldots,Y_{k+\ell}$ are any sections of $V$. Moreover, $\mathcal{J} = \{1,\ldots,k+\ell\}$ and $\mathbf{A} = (1,\ldots,k+\ell)$, see Convention
\ref{conv:mkxs}. The $\mathcal{J}$-multiindices have length $|\mathbf{A}| = k+\ell$, $|\mathbf{B}| = k$, $|\mathbf{C}| = \ell$. Also, $i$ is interior multiplication as in Convention \ref{conv:jjjjjutg}.\\
To check \eqref{rfgtr2}, use \eqref{eq:loiy2} with $Y = Y_{\mathbf{B}}$ and $z = Y_{\mathbf{C}}$ and apply it to
$\dmd\hskip-5.5pt/\hskip1pt u$. Then,
$$Y_{\mathbf{C}}\otimes \big(\dmd_{Y_{\mathbf{B}}} \dmd\hskip-5.5pt/\hskip1pt u\big)
- \dmd_{Y_{\mathbf{B}}} \big(Y_{\mathbf{C}}\otimes \dmd\hskip-5.5pt/\hskip1pt u\big)
 = - \big(\dmd_{Y_{\mathbf{B}}} Y_{\mathbf{C}}\big)\otimes \dmd\hskip-5.5pt/\hskip1pt u$$
Both sides are sections of $\Gamma(V^{\otimes \ell} \otimes {\wedge^{\ell} V^{\ast}} \otimes \mathcal{T}(V))$.
Contracting the first $\ell$ with the second $\ell$ factors, we obtain
(since diamonds commute with contractions)
$$i_{Y_{\mathbf{C}}} \big(\dmd_{Y_{\mathbf{B}}} \dmd\hskip-5.5pt/\hskip1pt u\big)
- \dmd_{Y_{\mathbf{B}}} \big(i_{Y_{\mathbf{C}}} \dmd\hskip-5.5pt/\hskip1pt u\big)
 = - i_{\dmd_{Y_{\mathbf{B}}} Y_{\mathbf{C}}} \big(\dmd\hskip-5.5pt/\hskip1pt u\big)$$
This is equivalent to (since $i_{Y_{\mathbf{C}}}i_{Y_{\mathbf{B}}} = i_{Y_{\mathbf{B}}\otimes Y_{\mathbf{C}}}$)
\begin{equation}\label{eq:krap}
i_{Y_{\mathbf{B}} \otimes Y_{\mathbf{C}}} \big(\dmd \dmd\hskip-5.5pt/\hskip1pt u\big)
= \dmd_{Y_{\mathbf{B}}} \big( \dmd\hskip-5.5pt/\hskip1pt_{Y_{\mathbf{C}}} u\big)
  -  \big(\dmd\hskip-5.5pt/\hskip1pt_{\dmd_{Y_{\mathbf{B}}} Y_{\mathbf{C}}}\big)\, u
\end{equation}
\end{remark}
\vskip 2mm

With Remark \ref{rem:jkjkjiu}, we obtain the following corollary of Proposition \ref{prop:ksaw2}.
\begin{corollary} \label{cor:loiya2}
For all $\dmd \in \conn^1$ and $Y_1,Y_2 \in \Gamma(V)$,
\begin{align*} 
\tfrac{1}{2}\sb{\dmd}{\dmd}_{Y_1\otimes Y_2} & = \big(i_{Y_1 \otimes Y_2} - i_{Y_2 \otimes Y_1}\big) \dmd\dmd\\
& = \dmd_{Y_1}\dmd_{Y_2} - \dmd_{Y_2}\dmd_{Y_1} - \dmd_{\dmd_{Y_1}Y_2 - \dmd_{Y_2}Y_1}
\end{align*}
\end{corollary}
\begin{definition} \label{def:lielielie} $\Lie(V,\cmet)$ is the subbundle of $\End(V)$ whose fiber at $p \in M$ is all $A \in \End(V)_p$
for which there is a $\lambda \in \R$ so that
\begin{equation}\label{eq:khsfdgjhgjsfd}
\cmetrep_p(AY_1,Y_2) + \cmetrep_p(Y_1,AY_2) = \lambda\, \cmetrep_p(Y_1,Y_2)
\end{equation}
for all $Y_1,Y_2\in V_p$. Here $\cmetrep_p \in (\Sym^2 V^{\ast})_p$ is a representative for $\cmet_p$. For each $k \geq 0$, set
$$\mathcal{R}^k = \Gamma\big( {\wedge^k V^{\ast}}\otimes \Lie(V,\cmet)\big)$$
\end{definition}
\begin{remark} The definition of the vector bundle $\Lie(V,\cmet)$ does not depend on the choice of a representative $\cmetrep$. The fibers of $\Lie(V,\cmet)$ have dimension 7. Each fiber is a Lie algebra isomorphic to the Lie algebra of the group $\R_+ \times O(1,3)$, the direct product of the multiplicative group of positive real numbers with the Lorentz group.
\end{remark}
\begin{proposition} For all $\dmd \in \conn^k_{\vertical}$ and $Y \in \Gamma(V^{\otimes k})$ and $Z \in \Gamma(V)$ set
$$\beta(\dmd)_Y Z = \dmd_YZ\; \in\; \Gamma(V)$$
Then $\beta(\dmd)_Y \in \Gamma(\Lie(V,\cmet)) \subset \Gamma(\End(V))$ and $\beta(\dmd) \in \mathcal{R}^k$. The map
\begin{align*}
\beta:\; \conn^k_{\vertical} & \to \mathcal{R}^k \\
\dmd & \mapsto \beta(\dmd)
\end{align*}
is a bijection.
\end{proposition}
\begin{proof} First, $\beta(\dmd) \in \Gamma({\wedge^k V^{\ast}}\otimes \End(V))$ because $\beta(\dmd)_YZ$ is linear over $C^{\infty}(M)$ in both $Y$ \emph{and} $Z$, by the assumption that $\dmd \in \conn^k_{\perp}$. We have to show that $\beta(\dmd) \in \mathcal{R}^k$. Let $\cmetrep$ be a representative of $\cmet$. Then
\begin{align*}
0 & = \dmd_Y(\cmetrep(Z_1,Z_2))\\
& = (\dmd_Y \cmetrep)(Z_1,Z_2) + \cmetrep(\dmd_Y Z_1,Z_2) + \cmetrep(Z_1,\dmd_YZ_2)\\
& = \mu(Y)\, \cmetrep(Z_1,Z_2) + \cmetrep(\beta(\dmd)_Y Z_1,Z_2) + \cmetrep(Z_1,\beta(\dmd)_YZ_2)
\end{align*}
for all $Z_1,Z_2 \in \Gamma(V)$, and $\mu$ as in (\ref{tre3}) of Definition \ref{def:mnnnbvc}. Hence, $\beta(\dmd) \in \mathcal{R}^k$. Also,
\begin{itemize}
\item $\beta$ is injective. In fact, $\beta(\dmd) = 0$ implies that $\dmd$ annihilates functions, sections of $V$ and, by equation \eqref{eq:jkjkjlfhkhf}, sections of $V^{\ast}$. By (a), (c) in Definition \ref{def:mnnnbvc}, we have $\dmd = 0$.
\item $\beta$ is surjective. Given $\Upsilon \in \mathcal{R}^k$, set
\begin{align*}
\dmd_Yf & = 0 & \dmd_YZ & = \Upsilon_YZ & 
(\dmd_Y\xi)(Z) & = - \xi(\Upsilon_YZ) 
\end{align*}
for all $f\in C^{\infty}(M)$, $Z \in \Gamma(V)$, $\xi \in \Gamma(V^{\ast})$ and all $Y \in \Gamma(V^{\otimes k})$. 
Together with (a),(c) in Definition \ref{def:mnnnbvc}, they uniquely determine $\dmd_Yu$ for all $u \in \Gamma(\mathcal{T}(V))$, and (b), (d), (e) in Definition \ref{def:mnnnbvc} are automatic. $\dmd \in \conn^k_{\vertical}$ satisfies $\beta(\dmd) = \Upsilon$.
\end{itemize}
\qed
\end{proof}

%% file: SectionLorentzian.tex
\section{From Diamonds of degree one to Lorentzian Geometry} \label{sec:kksasm}
In this section, we characterize the elements of $\conn^1$ that correspond to Lorentzian geometries. Conversely, we show that every Lorentzian manifold (locally) arises from an element of $\conn^1$.
The Einstein vacuum equations are reinterpreted as conditions on elements of $\conn^1$, to motivate their reformulation in Section \ref{sec:llllllllla21}.\\
This section is outside the overall technical development of this paper.
Its purpose is to connect the present formalism with traditional approaches.
\begin{proposition}\label{prop:orpl} For all $\dmd \in \conn^1$ there is a unique vector bundle homomorphism
\begin{equation}\label{eq:kiuz7}
\mathcal{E}^{\dmd}:\; V \to TM
\qquad \text{or, equivalently,}\qquad \mathcal{E}^{\dmd} \in \Gamma(V^{\ast}\otimes TM)
\end{equation}
such that $(\mathcal{E}^{\dmd}(Y))(f) = \dmd_Y(f)$ for all $Y \in \Gamma(V)$ and $f \in C^{\infty}(M)$.
\end{proposition}
\begin{proof}
The operator $\dmd_Y$ acts as a derivation on $C^{\infty}(M)$
and is linear over $C^{\infty}(M)$ in $Y$, by Definition \ref{def:mnnnbvc}.  \qed
\end{proof}
\begin{definition} \label{def:isome} A $\dmd \in \conn^1$ is called \emph{non-degenerate} if and only if $\mathcal{E}^{\dmd}$ is a vector bundle isomorphism.
The canonical extension of $\mathcal{E}^{\dmd}$ from $V$ to $\mathcal{T}(V)$ is also denoted by
$$\mathcal{E}^{\dmd}:\;\; \mathcal{T}(V) \to \mathcal{T}(TM)$$
The extension is a vector bundle isomorphism determined by
\begin{itemize}
\item $\mathcal{E}^{\dmd}(f) = f$ for all $f\in C^{\infty}(M)$
\item $\mathcal{E}^{\dmd}(u\otimes v) = \mathcal{E}^{\dmd}(u)\otimes \mathcal{E}^{\dmd}(v)$ for all $u,v \in \Gamma(\mathcal{T}(V))$
\item $\mathcal{E}^{\dmd}(I_V) = I_{TM}$ where $I_V \in \Gamma(\End(V))$, $I_{TM}\in \Gamma(\End(TM))$ are the identities
\end{itemize}
\end{definition}
\begin{proposition} \label{prop:nnnnnnabl} Let $\dmd \in \conn^1$ be non-degenerate. Let $\mathcal{E} = \mathcal{E}^{\dmd}$ and set
$$\nabla^{\dmd}: \Gamma(\mathcal{T}(TM)) \to \Gamma(T^{\ast}M\otimes \mathcal{T}(TM))
\qquad 
\nabla^{\dmd}_X u = \mathcal{E} \Big(\dmd_{\mathcal{E}^{-1}(X)} \mathcal{E}^{-1}(u)\Big)$$
for all $X \in \Gamma(TM)$ and $u\in \Gamma(\mathcal{T}(TM))$. Then $\nabla^{\dmd}$ is a connection on the tensor bundle $\mathcal{T}(TM)$ such that for all $X \in \Gamma(TM)$,
\begin{itemize}
\item $\nabla^{\dmd}$ is linear over $\R$
\item $\nabla_X^{\dmd}f = X(f)$ for all $f \in C^{\infty}(M)$
\item $\nabla^{\dmd}_X$ maps $C^{\infty}(M) \to C^{\infty}(M)$, $\Gamma(TM)\to \Gamma(TM)$ and $\Gamma(T^{\ast}M) \to \Gamma(T^{\ast}M)$
\item $\nabla_X^{\dmd}(u\otimes v) = (\nabla_X^{\dmd}u)\otimes v + u\otimes (\nabla_X^{\dmd}v)$
for all $u,v \in \Gamma(\mathcal{T}(TM))$
\item $\nabla^{\dmd}I = 0$ where $I \in \Gamma(\End(TM))$ is the identity.
\end{itemize}
\end{proposition}
\begin{proof} By direct verification. \qed
\end{proof}
\begin{lemma}\label{lem:juzy6} Let $\dmd \in \conn^1$ be non-degenerate. Let $\nabla = \nabla^{\dmd}$, $\mathcal{E} = \mathcal{E}^{\dmd}$. For all $X_i \in \Gamma(TM)$, $i=1,2$, and $v \in \Gamma(\mathcal{T}(TM))$ and corresponding
$Y_i = \mathcal{E}^{-1}(X_i) \in \Gamma(V)$, $i=1,2$, and $z = \mathcal{E}^{-1}(v) \in \Gamma(\mathcal{T}(V))$:
\begin{enumerate}[(a)]
\item $\big(\nabla_{X_1}\nabla_{X_2} - \nabla_{X_2}\nabla_{X_1} - \nabla_{\nabla_{X_1}X_2 - \nabla_{X_2}X_1}\big)v = \frac{1}{2}\, \mathcal{E}\big(\sb{\dmd}{\dmd}_{Y_1 \otimes Y_2} z\big)$
\item $\sb{\dmd}{\dmd} \in \conn^2_{\vertical}$ if and only if $\nabla$ is torsion-free
\end{enumerate}
Let $\cmetrep$ be a representative for $\cmet$ and let $\dmd \cmetrep = \mu \otimes \cmetrep$ as in (\ref{tre3}) of Definition \ref{def:mnnnbvc}. Let $\nu = \mathcal{E}(\mu) \in \Gamma(T^{\ast}M)$. 
For all $X_i$ and $Y_i$ as above, $i=1,2$, and all $f \in C^{\infty}(M)$:
\begin{enumerate}
\item[(c)] $\nabla_{X_1}\big(\mathcal{E}(e^f \cmetrep)\big) = e^f(\dd f + \nu)(X_1)\, \mathcal{E}(\cmetrep)$
\item[(d)] $\dd \nu(X_1,X_2)\,\cmetrep - \nu\Big(\nabla_{X_1}X_2 - \nabla_{X_2}X_1 - [X_1,X_2]\Big) \cmetrep = \tfrac{1}{2}\sb{\dmd}{\dmd}_{Y_1 \otimes Y_2}\cmetrep$
\end{enumerate}
\end{lemma}
\begin{proof} We verify (a) through (d):
\begin{enumerate}[(a)]
\item The left hand side is equal to $\mathcal{E}\big(\big({\dmd}_{Y_1} \dmd_{Y_2} - \dmd_{Y_2}\dmd_{Y_1} - \dmd_{\dmd_{Y_1}Y_2 - \dmd_{Y_2}Y_1}\big) z\big)$, 
by the definition of $\nabla = \nabla^{\dmd}$, see Proposition \ref{prop:nnnnnnabl}.
Now use Corollary \ref{cor:loiya2}. 
\item Let $v\in C^{\infty}(M)$ in (a). Then $z = v$.
We obtain $-\nabla_{T(X_1,X_2)} v = \tfrac{1}{2} \sb{\dmd}{\dmd}_{Y_1 \otimes Y_2} v$. The torsion $T$ of $\nabla$ vanishes if and only if $\sb{\dmd}{\dmd} \in \conn^2_{\vertical}$.
\item 
\begin{align*}
\nabla_{X_1} \big(\mathcal{E}(e^f\cmetrep)\big) & = e^f\,\dd f(X_1)\, \mathcal{E}(\cmetrep) + e^f \, \nabla_{X_1} \big(\mathcal{E}(\cmetrep)\big)\\
\nabla_{X_1} \big(\mathcal{E}(\cmetrep)\big) & = \mathcal{E}\big(\dmd_{\mathcal{E}^{-1}(X_1)} \cmetrep \big)
= \mathcal{E}\big(\mu(\mathcal{E}^{-1}(X_1))\, \cmetrep \big) = \nu(X_1)\, \mathcal{E}(\cmetrep)
\end{align*}
\item Let $z = \cmetrep$ in (a). Then $v = \mathcal{E}(\cmetrep)$. Rewrite the result using (c) with $f=0$.
\end{enumerate}
This concludes the proof. \qed
\end{proof}
\begin{convention} Let $\pi: E \to B$ be a vector bundle. For every $S \in \Gamma(\End(E))$ we denote by $\tr(S) \in C^{\infty}(B)$ its trace as a linear map.
\end{convention}
\begin{proposition}\label{prop:kkkkkkkkkkkkks}
Let $M$ be simply connected. Let $\dmd \in \conn^1$ and suppose
\begin{enumerate}[(a)]
\item\label{ccciii} $\dmd$ is non-degenerate
\item\label{cond:uzh} $\frac{1}{2}\sb{\dmd}{\dmd} \in \conn^2_{\vertical}$
\item\label{eq:l8j7h6g5f} $\tr (\Upsilon_{Y_1 \otimes Y_2}) = 0$ for all $Y_1,Y_2 \in \Gamma(V)$, where
$$\Upsilon = \beta(\tfrac{1}{2}\sb{\dmd}{\dmd}) \in \mathcal{R}^2$$
\end{enumerate}
Fix any representative $\cmetrep'$ for $\cmet$, and let $\dmd \cmetrep' = \mu \otimes \cmetrep'$ as in (\ref{tre3}) of Definition \ref{def:mnnnbvc}.\\
\emph{Part 1:} The 1-form $\nu = \mathcal{E}^{\dmd}(\mu) \in \Gamma(T^{\ast}M)$ is exact, $\nu = - \dd f$ with $f \in C^{\infty}(M)$.\\
\emph{Part 2:} Let $\cmetrep$ be a representative of $\cmet$. Then $\nabla^{\dmd}$ is the Levi-Civita connection for the Lorentzian metric $\mathcal{E}^{\dmd} (\cmetrep) \in \Gamma(\Sym^2 T^{\ast}M)$ if and only if $\cmetrep = e^{f+C} \cmetrep'$ for some $C \in \R$.\\
\emph{Part 3:} The associated Riemann curvature $R^{\dmd}$ is given by
\begin{equation}\label{eq:lokukloiuhjjhdzfht}
R^{\dmd}(X_1,X_2)X_3 = \mathcal{E}^{\dmd}\Big(\Upsilon_{Y_1 \otimes Y_2}Y_3\Big)
\;\; \in \;\; \Gamma(TM)
\end{equation}
for all $X_i \in \Gamma(TM)$ and $Y_i = (\mathcal{E}^{\dmd})^{-1}(X_i) \in \Gamma(V)$, $i=1,2,3$.
\end{proposition}
\begin{remark} \label{rem:afterkkkkkkks}
Part 2 of Proposition \ref{prop:kkkkkkkkkkkkks} implies that: \emph{There is a representative $\cmetrep$  of $\cmet$, unique up to an overall constant multiplicative factor, such that $\nabla^{\dmd}$ is the Levi-Civita connection for $\mathcal{E}^{\dmd}(\cmetrep)$.} In particular, the assignment $\dmd \mapsto \mathcal{E}^{\dmd}(\cmetrep)$ is canonical (independent of the choice of $\cmetrep'$), modulo an overall constant multiplicative factor.
\end{remark}
\begin{proof}
We use Lemma \ref{lem:juzy6} with the understanding that the representative for $\cmet$ in Lemma \ref{lem:juzy6} is $\cmetrep'$. Then $\nu$ in Lemma \ref{lem:juzy6} coincides with $\nu$ in Proposition \ref{prop:kkkkkkkkkkkkks}.\\
Part 1:  (\ref{cond:uzh}) implies that $\nabla^{\dmd}$ is torsion-free
by Lemma \ref{lem:juzy6}.(b).
Then $\dd \nu(X_1,X_2) \cmetrep' = \tfrac{1}{2}\sb{\dmd}{\dmd}_{Y_1\otimes Y_2}\cmetrep'$ by Lemma \ref{lem:juzy6}.(d).
Contracting with $(\cmetrep')^{-1}$ gives $4\,\dd \nu(X_1,X_2) = \tfrac{1}{2}\, i_{(\cmetrep')^{-1}} \big(\sb{\dmd}{\dmd}_{Y_1\otimes Y_2}\cmetrep'\big)$ where $i$ denotes interior multiplication. That is, both factors of  $\big(\sb{\dmd}{\dmd}_{Y_1\otimes Y_2}\cmetrep'\big) \in \Gamma(\Sym^2 V^{\ast})$
are contracted with $(\cmetrep')^{-1} \in \Gamma(\Sym^2 V)$.
Let $\mathcal{I}$, $F_{(a)}$ and $\lambda^{(a)}$, $a\in \mathcal{I}$, be as in Remark \ref{rem:jhkjhks}. Let $h_{ab}$ be the components of $\cmetrep'$, that is, $\cmetrep' =h_{ab} \lambda^{(a)}\otimes \lambda^{(b)}$ and $\cmetrep' = h^{ab} F_{(a)}\otimes F_{(b)}$, where $(h^{ab})$ is the inverse of $(h_{ab})$.
 By direct calculation, 
\begin{multline*}
\tfrac{1}{2}\, i_{(\cmetrep')^{-1}} \big(\sb{\dmd}{\dmd}_{Y_1\otimes Y_2}\cmetrep'\big)
= 
\big(\sb{\dmd}{\dmd}_{Y_1\otimes Y_2}\,\lambda^{(a)}\big)(F_{(a)})
= -  \lambda^{(a)} \big(\sb{\dmd}{\dmd}_{Y_1\otimes Y_2}\,F_{(a)}\big)\\
= - 2 \tr \big(\Upsilon_{Y_1\otimes Y_2}\big)
\end{multline*}
For the last equality, bear in mind that $\tfrac{1}{2}\sb{\dmd}{\dmd}$ and $\Upsilon$ coincide in their actions on sections of $V$. It follows from the last identity that $\dd \nu(X_1,X_2) = - \tfrac{1}{2}\tr \big(\Upsilon_{Y_1\otimes Y_2}\big)$, which vanishes for all $X_1,X_2 \in \Gamma(TM)$ by (c). Therefore, $\dd \nu = 0$. Since $M$ is simply connected, there is, by the Poincare Lemma, an $f\in C^{\infty}(M)$ with $\dd f = -\nu$.\\
Part 2: $\nabla^{\dmd}$ is torsion-free. $\nabla^{\dmd}$ is compatible with the Lorentzian metric $\mathcal{E}(e^F \cmetrep')$ if and only if
$F = f + C$ for some $C \in \R$, see Lemma \ref{lem:juzy6}.(c).\\
Part 3: Use Lemma \ref{lem:juzy6}.(a) with $v = X_3$ and recall that $\nabla^{\dmd}$ is torsion-free.
\qed
\end{proof}
\begin{remark}
To connect the Lie superalgebra identity $\sb{\dmd}{\sb{\dmd}{\dmd}} = 0$ with the classical algebraic and differential Bianchi identities for $R^{\dmd}$, we derive an identity. First of all, suppose that $\dmd \in \conn^1$ and $\slaa{\dmd} \in \conn^2_{\vertical}$. Then $\Upsilon = \beta(\slaa{\dmd}) \in \mathcal{R}^2$ is defined. Let $\mathcal{J} = \{1,2,3\}$.
For all $Y_i \in \Gamma(V)$, $i\in \mathcal{J}$, and $Z \in \Gamma(V)$, we have for any $\mathcal{J}$-multiindex  $\mathbf{C}$ with $|\mathbf{C}| = 2$,
$$i_{Y_{\mathbf{b}}\otimes Y_\mathbf{C} \otimes Z} \big(\dmd \Upsilon\big)
= i_{Y_{\mathbf{b}}\otimes Y_\mathbf{C}}\big(\dmd\slaa{\dmd} Z\big)
- i_{Y_\mathbf{C} \otimes Y_{\mathbf{b}}}\big(\slaa{\dmd} \dmd Z\big)
- \dmd_{\Upsilon_{Y_\mathbf{C}}Y_{\mathbf{b}}}Z
$$
Multiply by ${\antisym_{\mathbf{A}}}^{\mathbf{bC}}$, where $\mathbf{A} = (1,2,3)$, sum and obtain, by
equation \eqref{rfgtr1},
$${\antisym_{\mathbf{A}}}^{\mathbf{bC}} i_{Y_{\mathbf{b}}  \otimes Y_\mathbf{C}\otimes Z} \big(\dmd \Upsilon\big)
= \sb{\dmd}{\slaa{\dmd}}_{Y_{\mathbf{A}}} Z 
- {\antisym_{\mathbf{A}}}^{\mathbf{bC}} \dmd_{\Upsilon_{Y_\mathbf{C}}Y_{\mathbf{b}}}Z
$$
In the special case when 
$\sb{\dmd}{\dmd} \in \conn^2_{\vertical}$ and when $\slaa{\dmd} = \tfrac{1}{2}\sb{\dmd}{\dmd}$,
the first term on right hand side vanishes by the Lie superalgebra identity $\sb{\dmd}{\sb{\dmd}{\dmd}} = 0$. The left hand side is linear over $C^{\infty}(M)$ in $Z$, and so must be the right hand side. If $\dmd$ is non-degenerate, the last observation implies the "algebraic Bianchi identity"
$${\antisym_{\mathbf{A}}}^{\mathbf{bC}} \Upsilon_{Y_\mathbf{C}}Y_{\mathbf{b}} = 0$$
where
$\Upsilon =  \beta\big(\tfrac{1}{2}\sb{\dmd}{\dmd}\big)$.
Consequently, we also have the "differential Bianchi identity"
$${\antisym_{\mathbf{A}}}^{\mathbf{bC}} i_{Y_{\mathbf{b}} \otimes Y_\mathbf{C} \otimes Z} \big(\dmd \Upsilon\big) = 0$$
Finally, if $\dmd$ satisfies all the assumptions of Proposition \ref{prop:kkkkkkkkkkkkks}, then we obtain the traditional Bianchi identities for the associated Riemann curvature $R^{\dmd}$.
\end{remark}
\begin{proposition}\label{prop:kjxy}
Let $M$ be simply connected, and assume we are given
\begin{enumerate}[(a)]
\item\label{ewsdew2} a vector bundle isomorphism $\slaa{\mathcal{E}}: V \to TM$
\item\label{ewsdew1} a representative $\slaa{\cmetrep}$ for $\cmet$
\end{enumerate}
Let $\cmetrep' = \slaa{\cmetrep}$ in Proposition \ref{prop:kkkkkkkkkkkkks}. Then, there is a unique $\dmd \in \conn^1$ which satisfies the assumptions of Proposition \ref{prop:kkkkkkkkkkkkks} such that $\mathcal{E}^{\dmd} = \slaa{\mathcal{E}}$ and such that $\mu = 0$ in Proposition \ref{prop:kkkkkkkkkkkkks}.
\end{proposition}
\begin{remark}
Observe that (\ref{ewsdew2}) and (\ref{ewsdew1}) induce the Lorentzian metric $\slaa{\mathcal{E}}(\slaa{\cmetrep})$ on $M$. Conversely, every Lorentzian metric arises locally from such a construction.
\end{remark}
\begin{proof}
We use Lemma \ref{lem:juzy6} with the understanding that the representative for $\cmet$ in Lemma \ref{lem:juzy6} is $\slaa{\cmetrep}$. Then $\nu$ in Lemma \ref{lem:juzy6} coincides with $\nu$ in Proposition \ref{prop:kkkkkkkkkkkkks}.\\
We first prove existence.
The canonical extension of $\slaa{\mathcal{E}}$ from $V$ to $\mathcal{T}(V)$ is also denoted by
$\slaa{\mathcal{E}}:  \mathcal{T}(V)\to \mathcal{T}(TM)$ (just as in Definition \ref{def:isome}).
Let $\slab{\nabla}$ be the Levi-Civita connection associated with $\slaa{\mathcal{E}}(\slaa{\cmetrep})
\in \Gamma(\Sym^2 T^{\ast}M)$, a metric with signature $(-,+,+,+)$.
For all $Y \in \Gamma(V)$ and $u\in\Gamma(\mathcal{T}(V))$, set $\dmd_Y u = \slaa{\mathcal{E}}^{-1}(\slab{\nabla}_{\slaz{\mathcal{E}}(Y)}\slaa{\mathcal{E}}(u)) \in \Gamma(\mathcal{T}(V))$.
By direct inspection, $\dmd \in \conn^1$ (see Definition \ref{def:mnnnbvc}). 
Then $\slaa{\mathcal{E}} = \mathcal{E}^{\dmd}$ and $\slab{\nabla} = \nabla^{\dmd}$. In particular, $\dmd$ is non-degenerate. 
Lemma \ref{lem:juzy6}.(b) implies that $\sb{\dmd}{\dmd}\in \conn^2_{\vertical}$, because $\nabla^{\dmd} = \slab{\nabla} $ is torsion-free.
Lemma \ref{lem:juzy6}.(c) implies $\nu = 0$ because $\nabla^{\dmd} = \slab{\nabla}$ is compatible with the metric $ \slaa{\mathcal{E}}(\slaa{\cmetrep})$.
Now Lemma \ref{lem:juzy6}.(d) implies $\frac{1}{2}\sb{\dmd}{\dmd}_{Y_1\otimes Y_2} \slaa{\cmetrep} = 0$ for all $Y_1, Y_2$.
This implies $\tr(\Upsilon_{Y_1\otimes Y_2}) = 0$, where $\Upsilon = \beta(\frac{1}{2}\sb{\dmd}{\dmd})$. This concludes the existence proof. 
To prove uniqueness, assume there are two such $\dmd \in \conn^1$. Then their $\mathcal{E}^{\dmd} = \slaa{\mathcal{E}}$ coincide, and their $\nabla^{\dmd}$ coincide, because they are the Levi-Civita connection for the same metric $\mathcal{E}^{\dmd}(\slaa{\cmetrep})$ by Proposition \ref{prop:kkkkkkkkkkkkks}. Then the two $\dmd$'s must be the same. \qed
\end{proof}
\begin{proposition} \label{prop:1234}
Let $M$ be simply connected. Let $\dmd \in \conn^1$ be non-degenerate. The following are equivalent:
\begin{enumerate}[(a)]
\item $\dmd$ satisfies the assumptions of Proposition \ref{prop:kkkkkkkkkkkkks}, and the associated Lorentzian manifold is Ricci-flat
\item $\frac{1}{2}\sb{\dmd}{\dmd} \in \conn^2_{\vacuum}$
\item there is an $\slaa{\dmd} \in \conn^2_{\vacuum}$ such that $\slaa{\dmd} = \tfrac{1}{2}\sb{\dmd}{\dmd}$ and $\sb{\dmd}{\slaa{\dmd}} = 0$
\end{enumerate}
See Definition \ref{def:ovac} below for $\conn^2_{\vacuum}$.
\end{proposition}
\begin{proof} (c) implies (b), and conversely, (b) implies (c) by setting $\slaa{\dmd} = \tfrac{1}{2}\sb{\dmd}{\dmd}$ and using the super Jacobi identity to conclude that $\sb{\dmd}{\slaa{\dmd}} = \tfrac{1}{2} \sb{\dmd}{\sb{\dmd}{\dmd}} = 0$. The equivalence of (a) and (b) follows by comparing
for each row of the following table the corresponding condition/assumption in Proposition \ref{prop:kkkkkkkkkkkkks} and Definition \ref{def:ovac}:
\begin{center}
\begin{tabular}{c|c}
\emph{Proposition \ref{prop:kkkkkkkkkkkkks}} & \emph{Definition \ref{def:ovac} with $\slaa{\dmd} = \tfrac{1}{2}\sb{\dmd}{\dmd}$, $k=2$}\\
\hline
(b) & the assumption $\slaa{\dmd} \in \conn^2_{\vertical}$\\
alg. Bianchi identity for $R^{\dmd}$ and \eqref{eq:lokukloiuhjjhdzfht} & (a)\\
(c) & (b)\\
Ricci flatness and \eqref{eq:lokukloiuhjjhdzfht} & (c.2)\\
\end{tabular}
\end{center}
This concludes the proof. \qed
\end{proof}

%% file: SectionReform.tex
\section{Reformulation of the Einstein vacuum equations} \label{sec:llllllllla21}
In the next definition, the index set $\mathcal{J} = \{1,\ldots,k+1\}$ and $\mathbf{A} = (1,\ldots,k+1)$, and $\mathbf{B}$ is a $\mathcal{J}$-multiindex of length $|\mathbf{B}| = k$.
\begin{definition} \label{def:ovac}
 The "vacuum subspace" $\conn^k_{\vacuum}\subset \conn^k_{\vertical}$, $k=2,3,4$, is the set of all
$\dmd \in \conn^k_{\vertical}$ such that the associated $\Upsilon = \beta(\dmd) \in \mathcal{R}^k$  satisfies for all $Y_i \in \Gamma(V)$, $i \in \mathcal{J}$,
\begin{enumerate}
\item[(a)] 
${\antisym_{\mathbf{A}}}^{\mathbf{Bc}} \Upsilon_{Y_{\mathbf{B}}} Y_{\mathbf{c}} = 0$
\item[(b)] $\tr (\Upsilon_{Y_{\mathbf{B}}}) = 0$ for $\mathbf{B} = (1,\ldots,k)$
\item[(c.2)] for $k=2$: $\mathbf{C}(\Upsilon) = 0$ where $\mathbf{C}$ is the contraction operator for
the index pair $(2,4)$
\item[(c.3)] for $k=3$: $\mathbf{C}(\Upsilon \otimes \cmetrep^{-1}) = 0$ where $\mathbf{C}$ contracts $(1,5), (3,6)$ and $(4,7)$
\end{enumerate}
In (c.2) we regard $\Upsilon$ as a section of 
$(V^{\ast})^{\otimes 3} \otimes V \supset {\wedge^2 V^{\ast}}\otimes \Lie(V,\cmet)$.\\
In (c.3) we regard $\Upsilon \otimes \cmetrep^{-1}$ as a section of
$(V^{\ast})^{\otimes 4}\otimes V^{\otimes 3} \supset {\wedge^3 V^{\ast}}\otimes \Lie(V,\cmet)\otimes \Sym^2 V$.  Here, $\cmetrep$ is any representative of $\cmet$.  All contractions are natural pairings of $V$ with $V^{\ast}$.
\end{definition}
See Definition \ref{def:ppp9} and Proposition \ref{prop:tandu} for a discussion of $\conn^k_{\vacuum}$ in index notation.
\vskip 1mm
We now adopt verbatim, from Section \ref{sec:abstractidentity}, the definitions of $\mathcal{D}$,  $A_0$ and $A_1$, with the understanding that $L_0 = \conn_0$
and $L_1 = \conn_1$, see Proposition \ref{prop:ksaw2}.
In particular, for all $\fdmd = (\dmd,\slaa{\dmd}) \in \conn^1 \times \conn^2 \subset A_1$
and $\fdmd' = (\dmd',\slaa{\dmd}') \in \conn^2 \times \conn^3\subset A_0$, we have
\begin{subequations}\label{eq:ljflksdjlsdjs}
\begin{align}\label{eq:lofot}
\mathcal{D}_{\fdmd}\fdmd & = \Big(\slaa{\dmd} - \tfrac{1}{2} \sb{\dmd}{\dmd},\; \sb{\dmd}{\slaa{\dmd}}\Big) && \in  \conn^2 \times \conn^3 \subset A_0\\
\label{eq:lofot2} \mathcal{D}_{\fdmd} \fdmd' & = \Big(\slaa{\dmd}' - \sb{\dmd}{\dmd'\,},\;
\sb{\dmd}{\slaa{\dmd}'\,} + \sb{\dmd'}{\slaa{\dmd}}\Big) && \in \conn^3 \times \conn^4 \subset A_1
\end{align}
\end{subequations}
The Einstein vacuum equations are now reformulated as:
\begin{equation}\label{eq:hkjhkhkfds}
\text{Find $\fdmd \in \conn^1 \times \conn^2_{\vacuum}$ such that
$\mathcal{D}_{\fdmd}\fdmd = 0$.}
\end{equation}
\begin{remark}
Proposition \ref{prop:1234} justifies the expression 
 "\emph{reformulation} of the Einstein vacuum equations". Notice that, in contrast to Proposition \ref{prop:1234}, we do not require $\omega$ to be non-degenerate. Degenerate solutions may not be physically interesting in themselves. However, they can be used as a mathematical tool, to construct nearby non-degenerate solutions.
\end{remark}
We now derive algebraic and differential identities.
\begin{lemma}\label{lem:jjjys}
For all $(k,{\ell}) \in \{(1,2),(2,2),(1,3)\}$, all $\dmd \in \conn^k$ and all $\slaa{\dmd} \in \conn^{\ell}_{\vacuum}$, we have $\sb{\dmd}{\slaa{\dmd}} \in \conn^{k+{\ell}}_{\vacuum}$.
\end{lemma}
\begin{proof}
In this proof, the index set $\mathcal{J} = \{1,\ldots,k+\ell\}$ and $\mathbf{A} = (1,\ldots,k+\ell)$. 
First show that $\sb{\dmd}{\slaa{\dmd}}\in \conn^{k+{\ell}}_{\vertical}$. Equations \eqref{rfgtr2},
\eqref{eq:krap} and  Definition \ref{def:ovac}.(a) for $\slaa{\dmd} \in \conn^{\ell}_{\vacuum}$ imply
\begin{equation}\label{eq:kiolkiiujk}
\sb{\dmd}{\slaa{\dmd}}_{Y_{\mathbf{A}}}u = {\antisym_{\mathbf{A}}}^{\mathbf{BC}}  \bigg(
i_{Y_{\mathbf{B}}\otimes Y_{\mathbf{C}}}\big(\dmd\slaa{\dmd} u\big) - \slaa{\dmd}_{Y_{\mathbf{C}}} \big(\dmd_{Y_{\mathbf{B}}} u\big) \bigg)
\end{equation}
We have used that $\Upsilon^{\slaz{\dmd}} = \beta(\slaa{\dmd}) \in \mathcal{R}^{\ell}$ satisfies $\Upsilon^{\slaz{\dmd}}_YZ = \slaa{\dmd}_YZ $ for all $Y \in \Gamma(V^{\otimes \ell})$ and $Z \in \Gamma(V)$.
The assumption $\slaa{\dmd} \in \conn^{\ell}_{\vertical}$ implies that $\slaa{\dmd}f = 0$ for all $f\in C^{\infty}(M)$, and consequently by equation \eqref{eq:kiolkiiujk}, $\sb{\dmd}{\slaa{\dmd}} f = 0$
for all $f \in C^{\infty}(M)$. Therefore, $\sb{\dmd}{\slaa{\dmd}} \in \conn^{k+{\ell}}_{\vertical}$.

We can now define $\Upsilon^{\slaz{\dmd}} = \beta(\slaa{\dmd}) \in \mathcal{R}^{\ell}$ and $\Upsilon^{\sb{\dmd}{\slaz{\dmd}}} = \beta(\sb{\dmd}{\slaa{\dmd}}) \in \mathcal{R}^{k+\ell}$. Equation \eqref{eq:kiolkiiujk}
with $u = Z \in \Gamma(V)$ is equivalent to 
\begin{equation}\label{eq:kmnhgbvfc}
\Upsilon^{\sb{\dmd}{\slaz{\dmd}}}_{Y_{\mathbf{A}}} Z 
= 
{\antisym_{\mathbf{A}}}^{\mathbf{BC}}  i_{Y_{\mathbf{B}}\otimes Y_{\mathbf{C}}\otimes Z}\big(\dmd \Upsilon^{\slaz{\dmd}}\big)
\end{equation}
Here $\dmd \Upsilon^{\slaz{\dmd}}$ is a section of $(V^{\ast})^{\otimes (k+\ell+1)} \otimes V \supset {\wedge^k V^{\ast}}\otimes {\wedge^{\ell} V^{\ast}}\otimes \Lie(V,\cmet)$.
We now check that $\sb{\dmd}{\slaa{\dmd}} \in \conn^{k+{\ell}}_{\vacuum}$, by showing (a), (b) in Definition \ref{def:ovac} for $\sb{\dmd}{\slaa{\dmd}}$. When $(k,\ell) = (1,2)$ we also have to check (c.3).
\begin{itemize}
\item The totally antisymmetric part of the right hand side of equation \eqref{eq:kmnhgbvfc} with respect to $Y_1,\ldots,Y_{k+{\ell}},Z$ vanishes
by (a) for $\slaa{\dmd} \in \conn^{\ell}_{\vacuum}$. Therefore, (a) holds for $\sb{\dmd}{\slaa{\dmd}}$.
\item $\dmd_{Y_{\mathbf{A}}}$ commutes with natural contractions (pairings of $V$ with $V^{\ast}$). Therefore, (b) for $\slaa{\dmd} \in \conn^{\ell}_{\vacuum}$ and equation \eqref{eq:kmnhgbvfc}  imply (b) for $\sb{\dmd}{\slaa{\dmd}}$.
\end{itemize}
This concludes the proof when $(k,{\ell}) \in \{(2,2),(1,3)\}$. From here, $(k,{\ell}) = (1,2)$.
\begin{itemize}
\item We must show (c.3). We must show that $\mathbf{C}(\Upsilon^{\sb{\dmd}{\slaz{\dmd}}}\otimes \cmetrep^{-1}) = 0$, where $\mathbf{C}$ contracts the index-pairs $(1,5), (3,6), (4,7)$ (see the explanation at the end of Definition \ref{def:ovac}). By writing out the sum on the right hand side of \eqref{eq:kmnhgbvfc} (there are $|P(1,2)| = 3$ terms), we see that it suffices to show that the contractions
\begin{equation}\label{ztkiuzt}
(3,5), (2,6), (4,7)\quad  \text{or}\quad (2,5), (1,6), (4,7)
\quad \text{or} \quad (1,5), (3,6), (4,7)
\end{equation}
of $(\dmd\Upsilon^{\slaz{\dmd}})\otimes \cmetrep^{-1} \in \Gamma((V^{\ast})^{\otimes 4}\otimes V^{\otimes 3})$ all vanish. Recall that there is a $\mu \in \Gamma(V^{\ast})$ such that $\dmd\cmetrep = \mu \otimes \cmetrep$. Consequently, $\dmd(\cmetrep^{-1}) = - \mu \otimes (\cmetrep^{-1})$. By the Leibniz rule,
$$\big(\dmd \Upsilon^{\slaz{\dmd}}\big)\otimes (\cmetrep^{-1}) = \Big(\dmd + \mu \otimes \Big)\big(\Upsilon^{\slaz{\dmd}}\otimes \cmetrep^{-1}\big)$$
The contractions listed in \eqref{ztkiuzt} indeed vanish, because $\slaa{\dmd} \in \conn^2_{\vacuum}$.
In the first set of pairings, the contraction $(3,5)$ suffices. In the second, $(2,5)$ suffices.
In the third, $(3,6)$ and $(4,7)$ together suffice, by (b) and (c.2) for $\slaa{\dmd} \in \conn^2_{\vacuum}$.
\end{itemize}
This concludes the proof. \qed
\end{proof}
\begin{proposition} \label{prop:lllxsy}
For all $\fdmd  \in \conn^1\times \conn^2_{\vacuum}$ and all
$\fdmd' \in \conn^2 \times \conn^3_{\vacuum}$,
\begin{subequations}
\begin{align}
\label{eq:mn1} \mathcal{D}_{\fdmd}\fdmd \;& \in\; \conn^{2} \times \conn^{3}_{\vacuum}\\
\label{eq:mn2} \mathcal{D}_{\fdmd} \mathcal{D}_{\fdmd} \fdmd & = 0\\
\label{eq:mn3} \mathcal{D}_{\fdmd}\fdmd'  \; & \in\; \conn^3 \times \conn^4_{\vacuum}
\end{align}
\end{subequations}
\end{proposition}
\begin{proof}
Equations \eqref{eq:mn1} and \eqref{eq:mn3} follow from Lemma \ref{lem:jjjys}, equation \eqref{eq:mn2} follows from Proposition \ref{prop:hhhgd}. \qed
\end{proof}
\begin{remark}
Equations \eqref{eq:mn1} and \eqref{eq:mn2} are, respectively, algebraic and differential identities for the left hand side of the equation $\mathcal{D}_{\fdmd}\fdmd = 0$.
\end{remark}

%% file: SectionIndices.tex
\section{Components and Multiindices}\label{sec:coord}
In this section, the previous  constructions are made concrete by introducing local coordinates and components. For this purpose, fix
\begin{itemize}
\item an index set $\mathcal{I}$ with $|\mathcal{I}| = 4$
\item a constant symmetric matrix $(g_{ab})_{a,b \in \mathcal{I}}$ with signature $(-,+,+,+)$
\item an open set $U \subset M$
\item a coordinate diffeomorphism $\diffeo: U \to \mathcal{U}\subset \R^4,\; p \mapsto (\diffeo^{\mu}(p))_{\mu=1,2,3,4}$
\item a representative $\cmetrep$  of $\cmet$ over $U$
\item sections $F_{(a)}$ of $V$ over $U$, $a\in \mathcal{I}$, such that
$\cmetrep(F_{(a)},F_{(b)}) = g_{ab}$
\end{itemize}
\begin{convention} \label{conv:111} We denote by $(g^{ab})_{a,b \in \mathcal{I}}$ the inverse of $(g_{ab})_{a,b\in \mathcal{I}}$.
\end{convention}
\begin{convention} $(\lambda^{(a)})_{a\in \mathcal{I}}$ are the sections of $V^{\ast}$ over $U$ dual to $(F_{(a)})_{a\in \mathcal{I}}$.
\end{convention}
\begin{convention} \label{conv:113}
Standard Cartesian coordinates on $\mathcal{U}\subset \R^4$ are denoted $(x^{\mu})_{\mu=1,2,3,4}$.
\end{convention}
\begin{convention}\label{conv:jjjjxsa}
Small Latin indices take values in the index set $\mathcal{I}$.
Capital Latin indices are multiindices, that is, elements of $\mathcal{I}^k$ for some $k\geq 0$. For example, $A = (a_1\ldots a_k)$ where $a_1,\ldots,a_k \in \mathcal{I}$. The length of a multiindex will be denoted by $|A| = k$.
Moreover, ${\antisym_A}^{BC}$ is introduced just as in Definition \ref{def:hzttr}, with the understanding that ordinary Latin indices refer to the index set $\mathcal{J} = \mathcal{I}$.
\end{convention}
\begin{convention}
For any multiindex $A = (a_1\ldots a_k)$, write $F_{(A)} = F_{(a_1)} \otimes \cdots \otimes F_{(a_k)}$.
\end{convention}
\begin{definition}\label{def:ppp9}
$S^k$ is the real vector space of all $(\XX,\YY) = ({\XX_{A}}^{\mu},{\YY_{A m}}^n)$, where $A$ is an $\mathcal{I}$-multiindex of length $|A| = k$ and $m,n \in \mathcal{I}$ and $\mu = 1,2,3,4$, such that
\begin{enumerate}
\item[(a)] $\XX$,
 $\YY$ are totally antisymmetric in their first $k$ lower indices,
\item[(b)] ${\YY_{A m}}^{\ell}\met_{\ell n} + {\YY_{An}}^{\ell} \met_{\ell m} = \tfrac{1}{2}{\YY_{A\ell}}^{\ell} \met_{mn}$ where $|A| = k$
\end{enumerate}

The "vertical subspace" $S^k_{\vertical}$ is the set of all $(\XX,\YY) \in S^k$ such that
\begin{enumerate}
\item[(c)] $\XX = 0$
\end{enumerate}

The "vacuum subspace" $S^k_{\vacuum}$, $2 \leq k \leq 4$, is the set of all $(\XX,\YY) \in S^k_{\vertical}$ such that
\begin{enumerate}
\item[(d)] ${\antisym_A}^B {\YY_B}^{n} = 0$ where $|A| = |B| = k+1$
\item[(e)] ${\YY_{Am}}^{\ell}\met_{\ell n} + {\YY_{An}}^{\ell}\met_{\ell m} = 0$ where $|A| = k$
\item[(f.2)] for $k=2$: ${\YY_{anm}}^n = 0$
\item[(f.3)] for $k=3$: $g^{bm}{\YY_{abnm}}^n = 0$
\end{enumerate}
\end{definition}
\begin{remark} Property (e) in Definition \ref{def:ppp9} implies ${\YY_{An}}^n = 0$.
\end{remark}
\begin{remark} \label{rem:kjhkfd} We have $\dim_{\R} S^k = 11 {4 \choose k}$ and $\dim_{\R} S^k_{\vertical} = 7 {4 \choose k}$ and
\begin{align*}
\dim_{\R} S^2_{\vacuum} & = 10 &
\dim_{\R} S^3_{\vacuum} & = 16 & 
\dim_{\R} S^4_{\vacuum} & = 6
\end{align*}
\end{remark}
Let $\conn^k(U)$ be defined just as in Definition \ref{def:mnnnbvc}, with $U$ instead of $M$. Similarly for $\conn^k_{\vertical}(U)$ and $\conn^k_{\vacuum}(U)$.
\begin{proposition} \label{prop:tandu}
\emph{Part 1}: 
Let $\dmd \in \conn^k(U)$. Set
\begin{subequations} \label{eq:kkkl}
\begin{align}
\label{eq:kkkl2} {(\XX^{\dmd})_A}^{\mu} \circ \diffeo & = \dmd_{F_{(A)}} \diffeo^{\mu}\\
\label{eq:kkkl3} \big({(\YY^{\dmd})_{Am}}^n \circ \diffeo\big) \, F_{(n)} & = \dmd_{F_{(A)}} F_{(m)}
\end{align}
\end{subequations}
Then $(\XX^{\dmd},\YY^{\dmd}) \in C^{\infty}(\mathcal{U},S^k)$.\\
\emph{Part 2}: For all
$(\XX^{\dmd},\YY^{\dmd}) \in C^{\infty}(\mathcal{U},S^k)$ there is a unique $\dmd \in \conn^k(U)$ so that  \eqref{eq:kkkl} hold.\\
\emph{Part 3}: $(\XX^{\dmd},\YY^{\dmd}) \in C^{\infty}(\mathcal{U},S^k_{\vertical})$ if and only if $\dmd \in \conn^k_{\vertical}(U)$.\\
\emph{Part 4}: $(\XX^{\dmd},\YY^{\dmd}) \in C^{\infty}(\mathcal{U},S^k_{\vacuum})$ if and only if $\dmd \in \conn^k_{\vacuum}(U)$.
\end{proposition}
\begin{remark}
For all $\dmd \in \conn^k(U)$, equations \eqref{eq:kkkl} imply that for all $f \in C^{\infty}(U)$:
\begin{subequations} \label{eq:llllkoi}
\begin{align}
\label{eq:mnbvcxy6}
\dmd_{F_{(A)}} f & = \Big( {(\XX^{\dmd})_{A}}^{\mu}\,  \tfrac{\p}{\p x^{\mu}}\big(f\circ \diffeo^{-1}\big)\Big)\circ \diffeo\\
\label{eq:iuzt7}
\dmd_{F_{(A)}} \lambda^{(m)} & = - \big({(\YY^{\dmd})_{An}}^m \circ \diffeo\big)\, \lambda^{(n)} 
\end{align}
\end{subequations}

\end{remark}
\begin{proof}[Proposition \ref{prop:tandu}] Recall that
$\cmetrep = \met_{ab}\lambda^{(a)}\otimes \lambda^{(b)}$ is a representative for $\cmet$ over $U$. Part 1: 
Use $\dmd \cmetrep = \mu \otimes \cmetrep$, where $\mu \in \Gamma({\wedge^k V^{\ast}}|_U)$, substitute
$\cmetrep = g_{ab}\lambda^{(a)}\otimes \lambda^{(b)}$ and use the Leibniz rule to show that 
$({(\YY^{\dmd})_{Am}}^{\ell} \circ \diffeo) \met_{\ell n}+  ({(\YY^{\dmd})_{An}}^{\ell}\circ \diffeo) \met_{m\ell}= - \mu(F_{(A)})\, \met_{mn}$. 
Multiply with $g^{mn}$, sum and obtain $-\tfrac{1}{2}({(\YY^{\dmd})_{A\ell}}^{\ell} \circ \diffeo) = \mu(F_{(A)})$. This implies (b) in Definition \ref{def:ppp9}.
Part 2:  Equations \eqref{eq:kkkl3}, \eqref{eq:mnbvcxy6} and
\eqref{eq:iuzt7} together with (\ref{tre1}), (\ref{tre2}) in Definition \ref{def:mnnnbvc} determine $\dmd$ uniquely. Properties (\ref{trexy}), (\ref{trenew}), (\ref{tre3}) in Definition \ref{def:mnnnbvc} are then automatic. This proves existence. $\dmd$ is unique, because for every $\dmd \in \conn^k(U)$, the equations \eqref{eq:kkkl} imply 
\eqref{eq:mnbvcxy6}, \eqref{eq:iuzt7}. Part 3:
$\dmd \in \conn^k_{\vertical}(U)$ iff $\dmd f = 0$ for all $f \in C^{\infty}(U)$ iff $\XX^{\dmd} = 0$, by equation \eqref{eq:mnbvcxy6}. 
Part 4 follows from Definition \ref{def:ovac}.
\qed
\end{proof}

\begin{proposition}\label{prop:sbcomp}
Let $\dmd \in \conn^k(U)$, $\slaa{\dmd} \in \conn^{\ell}(U)$. The superbracket $\sb{\dmd}{\slaa{\dmd}} \in \conn^{k+{\ell}}(U)$ has the  components
\begin{align*}
{(\XX^{\sb{\dmd}{\slaz{\dmd}}})_{A}}^{\mu} & = {\antisym_A}^{BC}\Big(
{(\XX^{\dmd})_B}^{\nu}\tfrac{\p}{\p x^{\nu}} {(\XX^{\slaz{\dmd}})_C}^{\mu}
-  {(\XX^{\slaz{\dmd}})_C}^{\nu}\tfrac{\p}{\p x^{\nu}} {(\XX^{\dmd})_B}^{\mu}\Big)\\
& \qquad -  {\antisym_A}^{BcE} {(\YY^{\dmd})_{Bc}}^{\ell} {(\XX^{\slaz{\dmd}})_{\ell E}}^{\mu} +  {\antisym_A}^{bDC} {(\YY^{\slaz{\dmd}})_{Cb}}^{\ell} {(\XX^{\dmd})_{\ell D}}^{\mu}\\
\intertext{and}
{(\YY^{\sb{\dmd}{\slaz{\dmd}}})_{Am}}^n & =  {\antisym_A}^{BC}\Big( {(\XX^{\dmd})_B}^{\mu} \tfrac{\p}{\p x^{\mu}} {(\YY^{\slaz{\dmd}})_{Cm}}^n
- {(\XX^{\slaz{\dmd}})_C}^{\mu} \tfrac{\p}{\p x^{\mu}} {(\YY^{\dmd})_{Bm}}^n\Big) \\
& \qquad + {\antisym_A}^{BC}\Big({(\YY^{\slaz{\dmd}})_{Cm}}^{\ell} {(\YY^{\dmd})_{B\ell}}^n - {(\YY^{\dmd})_{Bm}}^{\ell} {(\YY^{\slaz{\dmd}})_{C\ell}}^n \Big)\\
& \qquad -  {\antisym_A}^{BcE} {(\YY^{\dmd})_{Bc}}^{\ell} {(\YY^{\slaz{\dmd}})_{\ell E m}}^n  +  {\antisym_A}^{bDC} {(\YY^{\slaz{\dmd}})_{Cb}}^{\ell} {(\YY^{\dmd})_{\ell D m}}^n 
\end{align*}
The multiindices have length
\begin{align*}
|A| & = k+{\ell} &
|B| & = k &
|C| & = {\ell} &
|D| & = k-1 & 
|E| & = {\ell}-1
\end{align*}
\end{proposition}
\begin{proof}
By direct calculation, using \eqref{rfgtr2} and Proposition \ref{prop:tandu}.
Equation \eqref{rfgtr2} with $Y_i = F_{(a_i)}$ and $A = (a_1\ldots a_{k+\ell})$ implies
\begin{align*}
\sb{\dmd}{\slaa{\dmd}}_{F_{(A)}}u & = 
 {\antisym_A}^{BC} \Big(
\dmd_{F_{(B)}} \big(\slaa{\dmd}_{F_{(C)}} u\big)
- \slaa{\dmd}_{F_{(C)}} \big(\dmd_{F_{(B)}} u \big)
\Big)\\
& \qquad - {\antisym_A}^{BcE} \slaa{\dmd}_{(\dmd_{F_{(B)}}F_{(c)}),F_{(E)}}u + {\antisym_A}^{bDC} \dmd_{(\slaz{\dmd}_{F_{(C)}}F_{(b)}),F_{(D)}}u\\
& = {\antisym_A}^{BC} \Big(
\dmd_{F_{(B)}} \big(\slaa{\dmd}_{F_{(C)}} u\big)
- \slaa{\dmd}_{F_{(C)}} \big(\dmd_{F_{(B)}} u \big)
\Big)\\
& \qquad
- {\antisym_A}^{BcE}
\big({(\YY^{\dmd})_{Bc}}^{\ell} \circ \diffeo \big) \slaa{\dmd}_{F_{(\ell)},F_{(E)}}u
\\ &  \qquad
+{\antisym_A}^{bDC}
\big({(\YY^{\slaz{\dmd}})_{Cb}}^{\ell} \circ \diffeo \big)
 \dmd_{F_{(\ell)},F_{(D)}}u
\end{align*}
To calculate $\XX^{\sb{\dmd}{\slaz{\dmd}}}$, set $u = \diffeo^{\mu}$
and use \eqref{eq:kkkl} and \eqref{eq:llllkoi} repeatedly.
To calculate $\YY^{\sb{\dmd}{\slaz{\dmd}}}$, set $u = F_{(m)}$. \qed
\end{proof}
Propositions \ref{prop:tandu} and \ref{prop:sbcomp} enable us to write down all the equations of Section \ref{sec:llllllllla21} explicitly. See Section \ref{sec:instrman}. 
\section{Covariance}\label{sec:covariance}
For this section, fix
\begin{itemize}
\item $M$, $V$, $\cmet$ just as at the beginning of Section \ref{sec:diamonds}
\item another such triple $\widetilde{M}$, $\widetilde{V}$, $\widetilde{\cmet}$
\item open subsets $U \subset M$ and $\widetilde{U} \subset \widetilde{M}$
\item a diffeomorphism $\psi: \widetilde{U} \to U$
\item a vector bundle isomorphism $\phi: \widetilde{W} = \widetilde{V}|_{\widetilde{U}} \to W = V|_U$ so that $\pi_{W} \circ \phi = \psi \circ \pi_{\widetilde{W}}$
\end{itemize}
We require that
\begin{itemize}
\item for each representative $\widetilde{\cmetrep}$ of $\widetilde{\cmet}$ over $\widetilde{U}$,  $\phi(\widetilde{\cmetrep}) \in \Gamma(\Sym^2 W^{\ast})$ is a representative for $\cmet$ over $U$.
\end{itemize}
\begin{convention} As always, there is a canonical extension of $\phi$ to a vector bundle isomorphism $\mathcal{T}(\widetilde{W}) \to \mathcal{T}(W)$, which we also denote as $\phi$. For every section $u \in \mathcal{T}(\widetilde{W})$ we denote by $\phi(u) = \phi \circ u \circ \psi^{-1}$ the corresponding section of $\mathcal{T}(W)$.
\end{convention}
Let $\conn^k(U)$ and $\conn^k(\widetilde{U})$ be defined just as in Definition
\ref{def:mnnnbvc}.
\begin{proposition} \label{prop:hkhkfds}
For all $\dmd \in \conn^k(U)$ and all $\widetilde{Y} \in \Gamma(\widetilde{W}^{\otimes k})$ and $\widetilde{u} \in \mathcal{T}(\widetilde{W})$, set 
\begin{equation}\label{eq:khkhkjhfd}
\widetilde{\dmd}_{\widetilde{Y}} \widetilde{u} = \phi^{-1}\big(\dmd_Yu\big)
\end{equation}
where $Y = \phi(\widetilde{Y}) \in \Gamma(W^{\otimes k})$ and $u = \phi(\widetilde{u}) \in \Gamma(\mathcal{T}(W))$.
Then $\widetilde{\dmd} \in \conn^k(\widetilde{U})$. The map $\conn^k(U) \to \conn^k(\widetilde{U}),\, \dmd \mapsto \widetilde{\dmd} = \phi^{-1}(\dmd)$
\begin{itemize}
\item is a bijection that maps $\conn^k_{\vertical}(U) \to \conn^k_{\vertical}(\widetilde{U})$ and $\conn^k_{\vacuum}(U) \to \conn^k_{\vacuum}(\widetilde{U})$,
\item 
$\sb{\phi^{-1}(\dmd_1)}{\phi^{-1}(\dmd_2)} = \phi^{-1}\big(\sb{\dmd_1}{\dmd_2}\big)$ for all
 $\dmd_1, \dmd_2 \in \conn^k(\widetilde{U})$.
\end{itemize}
\end{proposition}
\begin{proof} By construction. \qed
\end{proof}
We will now spell out the transformation law $\dmd \mapsto \widetilde{\dmd}$ (see Proposition \ref{prop:hkhkfds}) in components. For this purpose, we fix additional objects, as at the beginning of Section \ref{sec:coord}:
\begin{itemize}
\item $\mathcal{I}$ and $(g_{ab})$
\item $\rho: U \to \mathcal{U}\subset \R^4$ and $\cmetrep$ and $(F_{(a)})$
\item $\widetilde{\rho}: \widetilde{U}\to \widetilde{\mathcal{U}}\subset \R^4 $ and $\widetilde{\cmetrep}$ and $(\widetilde{F}_{(a)})$
\end{itemize}
Define
\begin{itemize}
\item $\chi: \widetilde{\mathcal{U}} \to \mathcal{U}$ by the following commuting diagram:
\begin{equation}\label{eq:lodhkhkdff}
\xymatrix{
\widetilde{W} \ar[r]^{\phi} \ar[d]_{\pi_{\widetilde{W}}} & W \ar[d]^{\pi_W}\\
\widetilde{U} 
 \ar[r]^{\psi} \ar[d]_{\widetilde{\diffeo}} & U
\ar[d]^{\diffeo}\\
\widetilde{\mathcal{U}} 
\ar[r]^{\chi} & \mathcal{U} 
}
\end{equation}
\item $\Omega: \widetilde{\mathcal{U}} \to (0,\infty)$ by
\begin{equation}\label{eq:kjhkfdsiuz}
\widetilde{\cmetrep} = \phi^{-1}(\cmetrep)\; (\Omega \circ \widetilde{\diffeo})^{-2}
\end{equation}
\item a matrix valued map $({\Lambda^a}_b)_{a,b\in\mathcal{I}}$ on $\widetilde{\mathcal{U}}$ by
\begin{equation}\label{eq:kjkhzkhskf}
\widetilde{F}_{(a)} = \phi^{-1}\big(F_{(b)}\big) \,\big({\Lambda^b}_a \circ \widetilde{\diffeo} \big)
\end{equation}
\item the components ${J_{\nu}}^{\mu} \in C^{\infty}(\widetilde{\mathcal{U}})$ of the inverse of the Jacobian of $\chi$ by
\begin{equation}\label{eq:jacob}
{J_{\nu}}^{\mu} = \Big(\tfrac{\p}{\p x^{\nu}} (\chi^{-1})^{\mu}\Big)\circ \chi
\qquad \text{or, equivalently,} \qquad
 \big(\tfrac{\p}{\p \widetilde{x}^{\nu}} \chi^{\alpha}\big) {J_{\alpha}}^{\mu} = {\delta_{\nu}}^{\mu}
\end{equation}
\end{itemize}
\begin{convention}\label{conv:khkhfds}
Standard Cartesian coordinates on $\mathcal{U}\subset \R^4$ and $\widetilde{\mathcal{U}} \subset \R^4$ are denoted $(x^{\mu})_{\mu = 1,2,3,4}$ and $(\widetilde{x}^{\mu})_{\mu=1,2,3,4}$ respectively.
\end{convention}
\begin{remark} Equations
\eqref{eq:kjhkfdsiuz}, \eqref{eq:kjkhzkhskf} and $\cmetrep(F_{(a)},F_{(b)}) = g_{ab}$, $\widetilde{\cmetrep}(\widetilde{F}_{(a)},\widetilde{F}_{(b)}) = g_{ab}$ imply
\begin{equation}\label{eq:khkhsdhak}
g_{ab} = g_{k\ell} \,\big(\tfrac{1}{\Omega}{\Lambda^k}_a\big)\,\big(\tfrac{1}{\Omega}{\Lambda^{\ell}}_b\big)
\end{equation}
on $\widetilde{\mathcal{U}}$. In other words, $(\tfrac{1}{\Omega} {\Lambda^a}_b)$ is a Lorentz transformation matrix.
\end{remark}
\begin{proposition}\label{prop:covAR} Let $\dmd \in \conn^k(U)$ and $\widetilde{\dmd} = \phi^{-1}(\dmd) \in \conn^k(\widetilde{U})$. Let $(\XX,\YY)$ and $(\widetilde{\XX},\widetilde{\YY})$ be the components of $\dmd$ and $\widetilde{\dmd}$, respectively, as in Proposition \ref{prop:tandu}.
(These components are functions on $\mathcal{U}$ and $\widetilde{\mathcal{U}}$.) We have on $\widetilde{\mathcal{U}}$
\begin{subequations} \label{eq:kiujhz}
\begin{align}
\label{eq:kiujhz7} {\widetilde{\XX}_A}^{\phantom{A}\mu} & = \big({\XX_B}^{\nu}\circ \chi\big)\,{\Lambda^B}_A\, {J_{\nu}}^{\mu}\\
\label{eq:kiujhz8} {\widetilde{\YY}_{Am}}^{\phantom{Am}n} &  = \tfrac{1}{\Omega^2}({\YY_{Bk}}^{\ell}\circ \chi) {\Lambda^B}_A {\Lambda^k}_m {\Lambda_{\ell}}^n + \tfrac{1}{\Omega^2} ({\XX_B}^{\nu} \circ \chi) {\Lambda^B}_A  {J_{\nu}}^{\mu} \big(\tfrac{\p }{\p \widetilde{x}^{\mu}} {\Lambda^{\ell}}_m\big) {\Lambda_{\ell}}^n
\end{align}
Here $A = (a_1\ldots a_k)$, $B = (b_1\ldots b_k)$, ${\Lambda^B}_A = {\Lambda^{b_1}}_{a_1}\cdots {\Lambda^{b_k}}_{a_k}$ and ${\Lambda_{\ell}}^n = g_{\ell a}{\Lambda^a}_b g^{bn}$.
\end{subequations}
\end{proposition}
\begin{remark} Set $\varphi = \chi^{-1}$ and ${K_{\nu}}^{\mu} = \frac{\p}{\p x^{\nu}}\varphi^{\mu}$ and $\Theta = \Omega \circ \varphi$
and ${\Delta^a}_b = {\Lambda^a}_b \circ \varphi$. Then \eqref{eq:kiujhz} is equivalent to
\begin{subequations} \label{eq:kiujhz88}
\begin{align}
\label{eq:kiujhz788} {\widetilde{\XX}_A}^{\phantom{A}\mu} \circ \varphi & = {\XX_B}^{\nu}\,{\Delta^B}_A\, {K_{\nu}}^{\mu}\\
\label{eq:kiujhz888} {\widetilde{\YY}_{Am}}^{\phantom{Am}n} \circ \varphi &  = \tfrac{1}{\Theta^2}\,{\YY_{Bk}}^{\ell} {\Delta^B}_A {\Delta^k}_m {\Delta_{\ell}}^n + \tfrac{1}{\Theta^2}\,{\XX_B}^{\nu} {\Delta^B}_A   \big(\tfrac{\p }{\p x^{\nu}} {\Delta^{\ell}}_m\big) {\Delta_{\ell}}^n
\end{align}
\end{subequations}
\end{remark}
\begin{proof}[Proposition \ref{prop:covAR}] 
Calculate
\begin{align*}
{\widetilde{\XX}_A}^{\phantom{A}\mu}\circ \widetilde{\diffeo} & = \widetilde{\dmd}_{\widetilde{F}_{(A)}} \widetilde{\rho}^{\mu} \\
& = \big({\Lambda^B}_A \circ \widetilde{\diffeo}\big)\, \widetilde{\dmd}_{\phi^{-1}(F_{(B)})} \Big((\chi^{-1})^{\mu}\circ \diffeo \circ \psi\Big)\\
& = \big({\Lambda^B}_A \circ \widetilde{\diffeo}\big)\, \Big(\dmd_{F_{(B)}} \big((\chi^{-1})^{\mu}\circ \diffeo\big)\Big)  \circ \psi \\
& = \big({\Lambda^B}_A \circ \widetilde{\diffeo}\big)\, \Big({\XX_B}^{\nu} \tfrac{\p}{\p x^{\nu}} (\chi^{-1})^{\mu})\Big)  \circ \diffeo \circ \psi 
\end{align*}
Compose with $\widetilde{\diffeo}^{-1}$ from the right, and obtain equation \eqref{eq:kiujhz7}.
To show \eqref{eq:kiujhz8}, use
\begin{align*}
\big({\widetilde{\YY}_{Am}}^{\phantom{Am}n} \circ \widetilde{\diffeo}\big) \widetilde{F}_{(n)} & = \widetilde{\dmd}_{\widetilde{F}_{(A)}} \widetilde{F}_{(m)}
\end{align*}
(see equation \eqref{eq:kkkl3}) and calculate
\begin{align*}
& \big(\widetilde{\YY}_{Am}^{\phantom{Am}n} \circ \widetilde{\diffeo}\big) \,\phi^{-1}(F_{(\ell)})\,\big( {\Lambda^{\ell}}_n \circ \widetilde{\diffeo}\big) \\
& = \big({\Lambda^B}_A \circ \widetilde{\diffeo}\big)\;\widetilde{\dmd}_{\phi^{-1}(F_{(B)})} \Big(\phi^{-1}(F_{(\ell)})\, ({\Lambda^{\ell}}_m \circ \widetilde{\diffeo})\Big)\\
& = \big({\Lambda^B}_A \circ \widetilde{\diffeo}\big)\;\Big\{
({\Lambda^{k}}_m \circ \widetilde{\diffeo})\;\widetilde{\dmd}_{\phi^{-1}(F_{(B)})} \phi^{-1}(F_{(k)})\, 
+
\phi^{-1}(F_{(\ell)})\; \widetilde{\dmd}_{\phi^{-1}(F_{(B)})} ({\Lambda^{\ell}}_m \circ \widetilde{\diffeo})
\Big\}
\\
& = \big({\Lambda^B}_A \circ \widetilde{\diffeo}\big) \Big\{
({\Lambda^{k}}_m \circ \widetilde{\diffeo})
\;\phi^{-1}\big(\dmd_{F_{(B)}} F_{(k)}\big)
\\
& \hskip 50mm +
\Big(\dmd_{F_{(B)}} \big({\Lambda^{\ell}}_m \circ \chi^{-1}\circ \diffeo \big)\Big)\circ \psi
\Big\}\phi^{-1}(F_{(\ell)})\\
& = \big({\Lambda^B}_A \circ \widetilde{\diffeo}\big) \Big\{
({\Lambda^{k}}_m \circ \widetilde{\diffeo})
\big({\YY_{Bk}}^{\ell} \circ \diffeo \circ \psi\big) 
\\
& \hskip 50mm +
\Big({\XX_B}^{\nu} \tfrac{\p}{\p x^{\nu}} \big({\Lambda^{\ell}}_m \circ \chi^{-1}\big)\Big) \circ \diffeo \circ \psi 
\Big\}\;\phi^{-1}(F_{(\ell)})\\
& = \big({\Lambda^B}_A \circ \widetilde{\diffeo}\big) \Big\{
({\Lambda^{k}}_m \circ \widetilde{\diffeo})
\big({\YY_{Bk}}^{\ell} \circ \diffeo \circ \psi\big) 
\\
& \hskip 36mm +
\Big(\big(\tfrac{\p}{\p \widetilde{x}^{\mu}}{\Lambda^{\ell}}_m \big) \circ \widetilde{\diffeo}\Big)\;
\Big({\XX_B}^{\nu} \tfrac{\p}{\p x^{\nu}} \big(\chi^{-1}\big)^{\mu}\Big) \circ \diffeo \circ \psi 
\Big\}\;\phi^{-1}(F_{(\ell)})
\end{align*}
From both sides, factor out $\phi^{-1}(F_{(\ell)})$, compose with $\widetilde{\diffeo}^{-1}$ from the right, and obtain \eqref{eq:kiujhz8}. \qed
\end{proof}

\section{Instruction manual} \label{sec:instrman}
The purpose of this section is to state, in a self-contained and ready-to-use manner, definitions and propositions that express the reformulated Einstein vacuum equations \eqref{eq:hkjhkhkfds}, in explicit coordinate/index notation on an open subset of $\R^4$.
\vskip 2mm
For this section, fix
\begin{itemize}
\item a simply connected open subset $\mathcal{U}\subset \R^4$
\item an index set $\mathcal{I}$ with $|\mathcal{I}| = 4$
\item a constant symmetric matrix $(\met_{ab})_{a,b\in \mathcal{I}}$ with signature $(-,+,+,+)$
\end{itemize}
\noindent
The {\bf statements} of all definitions and propositions in this section are completely self-contained and make no reference to previous sections. The proofs, on the other hand, rely on the previous sections. We consider 
Definition \ref{def:hzttr}, Conventions \ref{conv:111}, \ref{conv:113}, \ref{conv:jjjjxsa} and Definition \ref{def:ppp9}
as being part of this section.
\vskip 2mm
\noindent In the next proposition, $A$, $B$, $C$, $D$ are $\mathcal{I}$-multiindices with length
$$|A| = |B| = 2 \qquad |C| = 3 \qquad |D| = 4$$
\begin{proposition} \label{prop:khkjhds} \emph{Part 1:} For all $\fdmd = ((E,\Gamma),(0,W)) \in C^{\infty}(\mathcal{U},S^1 \times S^2_{\vacuum})$
set
\begin{subequations}\label{eq:ljhkhfshks}
\begin{align}
 {T_{A}}^{\mu} & = - {\mathbf{A}_A}^{bc}\Big(
{E_b}^{\nu}\tfrac{\p}{\p x^{\nu}} {E_c}^{\mu}
 - {\Gamma_{bc}}^{\ell} {E_{\ell }}^{\mu}\Big)\\
{U_{Am}}^n & = {W_{Am}}^n - {\mathbf{A}_A}^{bc}\Big( {E_b}^{\mu} \tfrac{\p}{\p x^{\mu}} {\Gamma_{cm}}^n
+ {\Gamma_{cm}}^{\ell} {\Gamma_{b\ell}}^n
- {\Gamma_{bc}}^{\ell} {\Gamma_{\ell m}}^n\Big)\\
\label{343ztr7}
 {V_{Cm}}^n & =  {\mathbf{A}_C}^{bA}
\Big( {E_b}^{\mu} \tfrac{\p}{\p x^{\mu}} {W_{Am}}^n
+ {\Gamma_{b\ell}}^n {W_{Am}}^{\ell}  - {\Gamma_{bm}}^{\ell} {W_{A\ell}}^n
-2 {\Gamma_{A}}^{\ell} {W_{\ell b m}}^n\Big)
\end{align}
\end{subequations}
Then $\fdmd' = ((T,U),(0,V))$ is in $C^{\infty}(\mathcal{U},S^2 \times S^3_{\vacuum})$. In other words, there is a map 
\begin{align}
C^{\infty}(\mathcal{U},S^1 \times S^2_{\vacuum})
 & \to C^{\infty}(\mathcal{U},S^2 \times S^3_{\vacuum})\\
\notag \fdmd & \mapsto \fdmd'
\end{align}
which we again write as $\mathcal{D}_{\fdmd}\fdmd = \fdmd'$.\\
\emph{Part 2:} For all 
\begin{align*}
\fdmd & = ((E,\Gamma),(0,W)) \in C^{\infty}(\mathcal{U},S^1 \times S^2_{\vacuum})\\
\fdmd' & = ((T,U),(0,V)) \in C^{\infty}(\mathcal{U},S^2 \times S^3_{\vacuum})
\end{align*}
not necessarily $\fdmd' = \mathcal{D}_{\fdmd}\fdmd$, set
\begin{subequations}
\begin{align}
 {{\mathfrak T}_C}^{\mu} & = {\mathbf{A}_C}^{bA}\Big(
- {E_b}^{\nu}\tfrac{\p}{\p x^{\nu}} {T_A}^{\mu}
+ {T_A}^{\nu}\tfrac{\p}{\p x^{\nu}} {E_b}^{\mu}
 + 2{\Gamma_A}^{\ell} {T_{\ell b}}^{\mu} - {U_{Ab}}^{\ell} {E_{\ell}}^{\mu}\Big)\\
 {{\mathfrak U}_{Cm}}^n & = {V_{Cm}}^n - {\mathbf{A}_C}^{bA}\Big( {E_b}^{\mu} \tfrac{\p}{\p x^{\mu}} {U_{Am}}^n
- {T_A}^{\mu} \tfrac{\p}{\p x^{\mu}} {\Gamma_{bm}}^n
+ {U_{Am}}^{\ell} {\Gamma_{b\ell}}^n \\
\notag & \hskip 35mm - {\Gamma_{bm}}^{\ell} {U_{A\ell}}^n 
-2 {\Gamma_A}^{\ell} {U_{\ell b m}}^n  + {U_{Ab}}^{\ell} {\Gamma_{\ell m}}^n \Big)\\
{{\mathfrak V}_{Dm}}^n & =  {\mathbf{A}_D}^{bC}\Big( {E_b}^{\mu} \tfrac{\p}{\p x^{\mu}} {V_{Cm}}^n + {V_{Cm}}^{\ell} {\Gamma_{b\ell}}^n - {\Gamma_{bm}}^{\ell} {V_{C\ell}}^n
+ 3\,{U_C}^{\ell} {W_{\ell b m}}^n \Big)\\
\notag   &  \quad + {\mathbf{A}_D}^{AB}\Big( {T_A}^{\mu} \tfrac{\p}{\p x^{\mu}} {W_{Bm}}^n + {W_{Bm}}^{\ell} {U_{A\ell}}^n - {U_{Am}}^{\ell} {W_{B\ell}}^n -2\,{\Gamma_A}^{\ell} {V_{\ell B m}}^n\Big)
\end{align}
\end{subequations}
Then $\fdmd'' = (({\mathfrak T},{\mathfrak U}),(0,{\mathfrak V}))$ is in $C^{\infty}(\mathcal{U},S^3 \times S^4_{\vacuum})$. In other words, there is a map
\begin{align}
C^{\infty}(\mathcal{U},S^1 \times S^2_{\vacuum}) \times C^{\infty}(\mathcal{U},S^2 \times S^3_{\vacuum}) 
& \to C^{\infty}(\mathcal{U},S^3 \times S^4_{\vacuum})\\
\notag (\fdmd,\fdmd') & \mapsto \fdmd''
\end{align}
which we again write as $\mathcal{D}_{\fdmd} \fdmd' = \fdmd''$.\\
\emph{Part 3:} For all $\fdmd \in C^{\infty}(\mathcal{U},S^1 \times S^2_{\vacuum})$,
\begin{equation} \label{eq:jjhkhd}
\mathcal{D}_{\fdmd}\mathcal{D}_{\fdmd} \fdmd = 0
\end{equation}
\end{proposition}
\begin{proof}
Warning, in this proof we consciously abuse notation, the symbols $U$ and $V$ are both given two meanings.\\
Let $K$ be the 4-dimensional real vector space spanned by elements $(k_{(a)})_{a\in \mathcal{I}}$. 
We use the previous sections, with the understanding that $M$, $V$, $\cmet$ at the beginning of Section \ref{sec:diamonds}
and $U$, $\diffeo$, $\cmetrep$, $F_{(a)}$ at the beginning of Section \ref{sec:coord} are:
\begin{itemize}
\item $M = \mathcal{U} \subset \R^4$ with trivial bundle $V = \mathcal{U} \times K$
\item $\diffeo: U = \mathcal{U} \to \mathcal{U}$ the identity transformation
\item $F_{(a)}: \mathcal{U}\ni x \mapsto (x,k_{(a)}) \in \mathcal{U}\times K$ constant sections
\item $\cmet$ is defined by declaring $\cmetrep$ to be a representative, where $\cmetrep(F_{(a)},F_{(b)}) = g_{ab}$.
\end{itemize}
Part 1: We identify $\fdmd =
((E,\Gamma),(0,W)) \in C^{\infty}(\mathcal{U},S^1 \times S^2_{\vacuum})$ with the corresponding 
$\fdmd \in \conn^1 \times \conn^2_{\vacuum}$, in the sense of Proposition \ref{prop:tandu}.
Let $\fdmd' = \mathcal{D}_{\fdmd}\fdmd \in \conn^2 \times \conn^3$ be given by equation \eqref{eq:lofot}. By Proposition \ref{prop:lllxsy}, $\fdmd' \in \conn^2 \times \conn^3_{\vacuum}$. Identify $\fdmd'$ with the corresponding $\fdmd' = ((T,U),(0,V)) \in C^{\infty}(\mathcal{U},S^2\times S^3_{\vacuum})$, in the sense of Proposition \ref{prop:tandu}. It follows from equation \eqref{eq:lofot} and Proposition \ref{prop:sbcomp} that $T,U,V$ are given by equations \eqref{eq:ljhkhfshks}.
For the last term in \eqref{343ztr7}, recall that $2 {\mathbf{A}_C}^{bA} {\Gamma_A}^{\ell}
= {\mathbf{A}_C}^{bmn}{\Gamma_{mn}}^{\ell}$, see Definition \ref{def:hzttr}.\\
Part 2: Analogous to Part 1, using equation \eqref{eq:lofot2}.\\
Part 3: This is now a corollary of Proposition \ref{prop:lllxsy}. \qed
\end{proof}
\vskip 2mm
The Einstein vacuum equations are reformulated as:
\begin{equation} \label{eq:lljlkjda}
\text{Find $\fdmd \in C^{\infty}(\mathcal{U},S^1 \times S^2_{\vacuum})$ such that
$\mathcal{D}_{\fdmd}\fdmd = 0$.}
\end{equation}
\begin{remark} \label{rem:hkjsdhkhfdkjhdskds}
By the proof of Proposition \ref{prop:khkjhds}, the coordinate construction of $\mathcal{D}$ and the abstract construction of $\mathcal{D}$ coincide. Therefore, \eqref{eq:hkjhkhkfds} and \eqref{eq:lljlkjda} are equivalent.
\end{remark}
\vskip 4mm
\begin{proposition} \label{prop:kjhhs} Suppose $\fdmd = ((E,\Gamma),(0,W)) \in C^{\infty}(\mathcal{U},S^1 \times S^2_{\vacuum})$ satisfies $$\mathcal{D}_{\fdmd}\fdmd = 0$$ and $({E_a}^{\mu})$ is invertible as a matrix at each point of $\mathcal{U}$, so that the four vector fields $E_a = {E_a}^{\mu}\frac{\p}{\p x^{\mu}}$, $a\in \mathcal{I}$, are a frame for each fiber of $T\mathcal{U}$.
\\
\emph{Part 1:} $\nu\in \Gamma(T^{\ast}\mathcal{U})$ given by $\nu(E_a) = -\tfrac{1}{2}{\Gamma_{an}}^{n}$ is exact, $\nu = - \dd f$ with $f \in C^{\infty}(\mathcal{U})$.\\
\emph{Part 2:} The Lorentzian metric $g$ on $\mathcal{U}$ given by
$g(E_a,E_b) = e^f g_{ab}$ has Levi-Civita connection  $\nabla_{E_a}E_m = {\Gamma_{am}}^n E_n$.\\
\emph{Part 3:} The associated Riemann curvature is given by $R(E_a,E_b)E_m = {W_{abm}}^{n}E_n$. In particular, the Ricci-curvature vanishes.
\end{proposition}
\begin{proof}
We adopt the conventions in the proof of Proposition \ref{prop:khkjhds}, up to and including the four bullets.
 We identify $\fdmd = ((E,\Gamma),(0,W))$ with the corresponding 
 $\fdmd = (\dmd,\slaa{\dmd}) \in \conn^1 \times \conn^2_{\vacuum}$, in the sense of Proposition \ref{prop:tandu}.
Recall Proposition \ref{prop:orpl}, Definition \ref{def:isome} and \eqref{eq:mnbvcxy6}. 
Observe that 
\begin{itemize}
\item $\dmd \in \conn^1$ is non-degenerate, because $({E_a}^{\mu})$ is invertible, and $\mathcal{E}^{\dmd}(F_{(a)}) = E_{a}$. In fact, for every $q \in C^{\infty}(\mathcal{U})$ we have $E_a(q) = {E_a}^{\mu}\tfrac{\p}{\p x^{\mu}} q = \dmd_{F_{(a)}} q$.
\item (a) in Proposition \ref{prop:1234} holds, because $M = \mathcal{U}$ is simply connected, $\dmd \in \conn^1$ is non-degenerate, (c) in Proposition \ref{prop:1234} holds by $\mathcal{D}_{\fdmd}\fdmd = 0$, and (c) implies (a).
\end{itemize}
Let $\cmetrep'$ in Proposition \ref{prop:kkkkkkkkkkkkks} be given by $\cmetrep'(F_{(a)},F_{(b)}) = g_{ab}$, and let $\dmd \cmetrep' = \mu \otimes \cmetrep'$. Then the 1-form $\nu \in \Gamma(T^{\ast}\mathcal{U})$ in Proposition \ref{prop:kkkkkkkkkkkkks} is given by $\nu(E_{a}) = \mu(F_{(a)}) = -\tfrac{1}{2}{\Gamma_{an}}^n$. For the last equality, use $\dmd_{F_{(a)}} \cmetrep' = \mu(F_{(a)}) \cmetrep'$ and equations \eqref {eq:kkkl} and \eqref{eq:llllkoi}.\\
We apply Parts 1, 2, 3 of Proposition \ref{prop:kkkkkkkkkkkkks}. $\nu = -\dd f$ with $f \in C^{\infty}(\mathcal{U})$.  $\nabla^{\dmd}$ is the Levi-Civita connection of $g = \mathcal{E}^{\dmd}(e^f \cmetrep')
= e^f \mathcal{E}^{\dmd}(\cmetrep')$, and $g(E_a,E_b) = e^f g_{ab}$. The connection $\nabla^{\dmd}_{E_a} E_m = \mathcal{E}^{\dmd}(\dmd_{F_{(a)}} F_{(m)}) 
= \mathcal{E}^{\dmd}({\Gamma_{am}}^nF_{(n)}) = {\Gamma_{am}}^n E_n$. The Riemann curvature is
$R(E_a,E_b)E_m = \tfrac{1}{2} \mathcal{E}^{\dmd}(\sb{\dmd}{\dmd}_{F_{(a)} \otimes F_{(b)}} F_{(m)})
= \mathcal{E}^{\dmd}(\slaa{\dmd}_{F_{(a)} \otimes F_{(b)}} F_{(m)})
= \mathcal{E}^{\dmd}({W_{abm}}^n F_{(n)}) = {W_{abm}}^{n} E_{n}$.
The Ricci-curvature vanishes by ${W_{anm}}^n = 0$. \qed
\end{proof}
\begin{proposition} \label{prop:lhjlkhksfdhkds}
Suppose $(E_a)_{a\in \mathcal{I}}$ is a frame for each fiber of $T\mathcal{U}$ and the Lorentzian metric $g$ given by $g(E_a,E_b) = g_{ab}$ is Ricci-flat. Then $g$ arises from a solution to $$\mathcal{D}_{\fdmd}\fdmd = 0$$ as in Proposition
\ref{prop:kjhhs}.
\end{proposition}
\begin{proof}
We adopt the conventions in the proof of Proposition \ref{prop:khkjhds}, up to and including the four bullets.
Define a vector bundle isomorphism
$\slaa{\mathcal{E}}: V \to T\mathcal{U}$ by $\slaa{\mathcal{E}}(F_{(a)}) = E_a$. Let $\slaa{\cmetrep}$ be given by $\slaa{\cmetrep}(F_{(a)},F_{(b)}) = g_{ab}$.  It is a representative of $\cmet$.
Then
$\slaa{\mathcal{E}}(\slaa{\cmetrep})(E_a,E_b) = \slaa{\cmetrep}(F_{(a)},F_{(b)}) = g_{ab} = g(E_a,E_b)$,  that is, $g = \slaa{\mathcal{E}}(\slaa{\cmetrep})$. 
Let $\dmd \in \conn^1$ be as in Proposition \ref{prop:kjxy}. Then $\dmd$ satisfies the assumptions of Proposition \ref{prop:kkkkkkkkkkkkks}, $\mathcal{E}^{\dmd} = \slaa{\mathcal{E}}$,
$\dmd \slaa{\cmetrep} = 0$, 
and the Lorentzian metric associated with $\dmd$
(see Remark \ref{rem:afterkkkkkkks}) is $\slaa{\mathcal{E}}(\slaa{\cmetrep}) = g$, which by assumption is Ricci-flat. By $\text{(a)} \Longrightarrow \text{(c)}$ in  Proposition \ref{prop:1234} (recall that $\mathcal{U}$ is simply connected) there is a $\slaa{\dmd} \in \conn^2_{\vacuum}$ so that $\fdmd = (\dmd,\slaa{\dmd})$ satisfies $\mathcal{D}_{\fdmd} \fdmd = 0$. Identify $\fdmd \in \conn^1\times \conn^2_{\vacuum}$ with the corresponding
$\fdmd = ((\XX^{\dmd},\YY^{\dmd}),(0,\YY^{\slaz{\dmd}})) \in C^{\infty}(\mathcal{U},S^1\times S^2_{\vacuum})$,  in the sense of Proposition \ref{prop:tandu}.
Then ${(\XX^{\dmd})_a}^{\mu}\frac{\p}{\p x^{\mu}} = \mathcal{E}^{\dmd}(F_{(a)}) = \slaa{\mathcal{E}}(F_{(a)}) = E_a$, that is, ${(\XX^{\dmd})_a}^{\mu} = {E_a}^{\mu}$.
Moreover, $\nu = 0$ in Proposition \ref{prop:kjhhs}
and we can choose $f = 0$.
Then the $g$'s in Proposition \ref{prop:kjhhs} and \ref{prop:lhjlkhksfdhkds} coincide.
 \qed
\end{proof}
\begin{proposition} \label{proplkhjkhsdk}
Let $\widetilde{\mathcal{U}}\subset \R^4$ be open. We use Convention \ref{conv:khkhfds}. Let $(\chi,\Lambda)$ be a pair,
\begin{itemize}
\item $\chi: \widetilde{\mathcal{U}} \to \mathcal{U}$ a diffeomorphism
\item $({\Lambda^a}_b)_{a,b\in\mathcal{I}} = \Omega\, ({L^a}_b)_{a,b\in\mathcal{I}}$ where $\Omega: \widetilde{\mathcal{U}} \to (0,\infty)$ and $({L^a}_b)_{a,b\in \mathcal{I}}$ is a matrix valued map on $\widetilde{\mathcal{U}}$ such that  $g_{ab} = g_{k\ell}{L^k}_a{L^{\ell}}_b$. 
\end{itemize}
and let
\begin{itemize}
\item ${J_{\nu}}^{\mu}$ be given by \eqref{eq:jacob}
\end{itemize}
To each $\dmd = (\XX,\YY) \in C^{\infty}(\mathcal{U},S^k)$ we associate $ \widetilde{\dmd} = (\widetilde{\XX},\widetilde{\YY}) \in C^{\infty}(\widetilde{\mathcal{U}},S^k)$ by
equations \eqref{eq:kiujhz}, or, equivalently, \eqref{eq:kiujhz88}. To
each $\fdmd = (\dmd,\slaa{\dmd}) \in C^{\infty}(\mathcal{U},S^k\times S^{k+1})$ we associate
$ \widetilde{\fdmd} = (\widetilde{\dmd},\widetilde{\slaa{\dmd}}) \in C^{\infty}(\widetilde{\mathcal{U}},S^k\times S^{k+1})$. Then:\\
\emph{Part 1:} For all $\dmd \in C^{\infty}(\mathcal{U},S^k)$,
$\fdmd \in C^{\infty}(\mathcal{U},S^1\times S^2_{\vacuum})$
and
$\fdmd' \in C^{\infty}(\mathcal{U},S^2\times S^3_{\vacuum})$:
\begin{enumerate}[(a)]
\item $\dmd \in C^{\infty}(\mathcal{U},S^k_{\vertical})$ if and only if
$\widetilde{\dmd} \in C^{\infty}(\widetilde{\mathcal{U}},S^k_{\vertical})$
\item $\dmd \in C^{\infty}(\mathcal{U},S^k_{\vacuum})$ if and only if $\widetilde{\dmd} \in C^{\infty}(\widetilde{\mathcal{U}},S^k_{\vacuum})$
\item $\widetilde{\fdmd}  \in C^{\infty}(\widetilde{\mathcal{U}},
S^1\times S^2_{\vacuum})$ and 
$\mathcal{D}_{\widetilde{\fdmd}} \widetilde{\fdmd} = \widetilde{\mathcal{D}_{\fdmd}\fdmd}$
\item $\widetilde{\fdmd}'  \in C^{\infty}(\widetilde{\mathcal{U}},
S^2\times S^3_{\vacuum})$ and 
$\mathcal{D}_{\widetilde{\fdmd}} \widetilde{\fdmd}' = \widetilde{\mathcal{D}_{\fdmd}\fdmd'}$
\end{enumerate}
Especially, $\widetilde{\fdmd}$ is a solution to \eqref{eq:lljlkjda} on $\widetilde{\mathcal{U}}$ if and only if $\fdmd$ is a solution to \eqref{eq:lljlkjda} on $\mathcal{U}$.\\
\emph{Part 2:} The composition of $(\chi,\Lambda)$ and $(\widetilde{\chi},\widetilde{\Lambda})$, where $\widetilde{\chi}: \widetilde{\widetilde{\mathcal{U}}}\to \widetilde{\mathcal{U}}$ and $\widetilde{\Lambda}$ is defined on $\widetilde{\mathcal{U}}$, is given by $(\chi \circ \widetilde{\chi}, (\Lambda \circ \widetilde{\chi}) \widetilde{\Lambda})$.
The inverse to $(\chi,\Lambda)$ is $(\chi^{-1},\Lambda^{-1} \circ \chi^{-1})$.
\end{proposition}
\begin{proof}
We adopt the conventions in the proof of Proposition \ref{prop:khkjhds}, up to and including the four bullets.
We make the same conventions for all quantities with tildes. We use Section \ref{sec:covariance}, with the understanding that the diffeomorphism $\psi: \widetilde{U} = \widetilde{\mathcal{U}} \to U = \mathcal{U}$ is given by $\psi = \chi$, and the vector bundle isomorphism $\phi: \widetilde{\mathcal{U}}\times \widetilde{K} \to \mathcal{U}\times K$ maps
$(\widetilde{x},\widetilde{k}_{(a)})$ to $(\chi(\widetilde{x}), k_{(b)}{\Lambda^b}_a(\widetilde{x}))$.
With these definitions, $\chi$, $\Omega$, ${\Lambda^a}_b$, ${J_{\nu}}^{\mu}$ as defined in Section \ref{sec:covariance} coincide with $\chi$, $\Omega$, ${\Lambda^a}_b$, ${J_{\nu}}^{\mu}$ in Proposition \ref{proplkhjkhsdk}. In other words, the diagram \eqref{eq:lodhkhkdff} commutes, and equations \eqref{eq:kjhkfdsiuz}, \eqref{eq:kjkhzkhskf}, \eqref{eq:jacob} hold. 
\begin{itemize}
\item We verify equation \eqref{eq:kjkhzkhskf}: For every $\widetilde{x} \in \widetilde{\mathcal{U}}$,
$$\big(\phi \circ \widetilde{F}_{(a)}\big)_{\widetilde{x}} = \phi(\widetilde{x},\widetilde{k}_{(a)})
= (\chi(\widetilde{x}),k_{(b)}{\Lambda^b}_a(\widetilde{x}))
= (F_{(b)}\circ \chi)_{\widetilde{x}}\, {\Lambda^b}_a(\widetilde{x})$$
That is, $\phi \circ \widetilde{F}_{(a)} = (F_{(b)} \circ \chi)\, {\Lambda^b}_a$. Compose with $\phi^{-1}$ from the left to obtain \eqref{eq:kjkhzkhskf}.
\item We verify equation \eqref{eq:kjhkfdsiuz}:
\begin{multline*}
\phi^{-1}\big(\cmetrep)(\widetilde{F}_{(a)},\widetilde{F}_{(b)}\big)
= \cmetrep\big(\phi(\widetilde{F}_{(a)}),\phi(\widetilde{F}_{(b)})\big) \circ \chi
=  \big(\cmetrep(F_{(k)},F_{(\ell)}) \circ \chi\big)\; {\Lambda^k}_a\, {\Lambda^{\ell}}_b\\
= g_{k\ell} {\Lambda^k}_a {\Lambda^{\ell}}_b =  \Omega^2 g_{ab}  = \Omega^2 \widetilde{\cmetrep}\big(\widetilde{F}_{(a)},\widetilde{F}_{(b)}\big)
\end{multline*}
\end{itemize}
We identify abstract diamonds and their components, in the sense of Proposition \ref{prop:tandu}. With this understanding, the maps
\begin{itemize}
\item $C^{\infty}(\mathcal{U},S^k) \to C^{\infty}(\widetilde{\mathcal{U}},S^k)$, $\dmd \mapsto \widetilde{\dmd}$ in Proposition \ref{proplkhjkhsdk}
\item $\conn^k(U) \to \conn^k(\widetilde{U})$, $\dmd \mapsto \phi^{-1}(\dmd)$ in Proposition \ref{prop:hkhkfds}
\end{itemize}
coincide, by Proposition \ref{prop:covAR}. Part 1: Now (a), (b) follow from Proposition
\ref{prop:tandu} and Proposition \ref{prop:hkhkfds}. The first statements in (c) and (d) follow from (a) and (b). The second statements in (c) and (d) follow from Remark \ref{rem:hkjsdhkhfdkjhdskds}, equations \eqref{eq:ljflksdjlsdjs}
and the fact that the map $\dmd \mapsto \widetilde{\dmd}$ commutes with the Lie superbracket, see Proposition \ref{prop:hkhkfds}. Part 2: Let $\psi$, $\phi$ and $\widetilde{\psi}$, $\widetilde{\phi}$ be the diffeomorphism and vector bundle isomorphism corresponding to the pairs $(\chi,\Lambda)$ and $(\widetilde{\chi},\widetilde{\Lambda})$. Then the pair $(\chi \circ \widetilde{\chi}, (\Lambda \circ \widetilde{\chi}) \widetilde{\Lambda})$ corresponds to $\psi \circ \widetilde{\psi}$, $\phi \circ \widetilde{\phi}$. Now, Part 2 follows from Proposition \ref{prop:hkhkfds}.
\qed
\end{proof}
We conclude this section with a few remarks:
\begin{remark} The (coordinate) first order differential operators $\mathcal{D}$ in Part 1 and Part 2 of Proposition 
\ref{prop:khkjhds} are classically defined when $\fdmd$ and $\fdmd'$ are of class $C^1$.
Especially, the left hand side of the Einstein vacuum equation $\mathcal{D}_{\fdmd} \fdmd = 0$ is well defined for any $\fdmd$ of class $C^1$.
 By continuity, the differential identity \eqref{eq:jjhkhd} holds for every $\fdmd$ of class $C^2$.
\end{remark}
\begin{remark}\label{rem:kjhkhkd}
It is essential to observe that there is a canonical subformalism of the formalism of this paper,
which informally speaking is obtained by putting all the $\mu \in \Gamma(V^{\ast})$ and $\nu \in \Gamma(T^{\ast}M)$ to  zero. More precisely, at the beginning of Section \ref{sec:diamonds}, we choose a section $\cmetrep_0 = \Gamma(\Sym^2 V^{\ast})$ with signature $(-,+,+,+)$ instead of a conformal section $\cmet$.
From this point on, every representative of $\cmet$ is replaced by $\cmetrep_0$. Then, in (e) of Definition 
\ref{def:mnnnbvc} we also require that $\mu = 0$.
Definition \ref{def:lielielie} for $\Lie(V,\cmet)$ is replaced by a Definition of $\Lie(V,\cmetrep_0)$ by putting $\lambda = 0$ in \eqref{eq:khsfdgjhgjsfd}, giving a vector bundle with fibers of dimension 6, and the definition of $\mathcal{R}^k$ is changed accordingly. In this subformalism, condition (c)
in Proposition \ref{prop:kkkkkkkkkkkkks} is vacuous. Definition \ref{def:ppp9}.(b) is replaced by 
${\YY_{Am}}^{\ell} g_{\ell n} + {\YY_{An}}^{\ell} g_{\ell m} = 0$. (The new condition differs from the old condition by ${\YY_{A\ell}}^{\ell} = 0$.)
Remark \ref{rem:kjhkfd} is replaced by $\dim_{\R}S^k = 10{4 \choose k}$ and $\dim_{\R}S^k_{\vertical} = 6{4 \choose k}$, while $\dim_{\R}S^k_{\vacuum}$ is unchanged. Now $\Omega\equiv 1$ in \eqref{eq:kjhkfdsiuz}. We emphasize the consequences that the subformalism has for Proposition \ref{prop:khkjhds}. In Part 1, we have the new condition ${\Gamma_{am}}^{m} = 0$, and the new conclusion ${U_{Am}}^m = 0$. In Part 2, we have the new conditions ${\Gamma_{am}}^m = 0$ and ${U_{Am}}^m = 0$, and the new conclusion ${{\mathfrak U}_{Cm}}^m = 0$.
Finally, 
Propositions 
\ref{prop:kjhhs} and \ref{prop:lhjlkhksfdhkds} as well as all the other propositions hold for the subformalism, with the understanding that in Proposition \ref{proplkhjkhsdk}, $\Omega \equiv 1$.
\end{remark}
\begin{remark}
The discussion of Appendix B to \cite{RT} is a precursor to the formalism of this paper, more precisely to the subformalism elaborated on in Remark \ref{rem:kjhkhkd}. To compare the two developments, one must be aware that:
\begin{itemize}
\item The ordering of the indices may differ.
\item Combinatorial factors may differ.
\item In contrast to \cite{RT}, indices are neither raised nor lowered in this paper.
\end{itemize}
\end{remark}

%% file: Paper.bbl
\begin{thebibliography}{}
%
%
\bibitem[Fr]{Fr}
Friedrich, H., Proc. R. Soc. Lond. A \textbf{375}, (1981) 169-184
\bibitem[NP]{NP}
Newman, E. and Penrose, R., J. Math. Phys. \textbf{3}, (1962) 566-578
\bibitem[Sh]{Sh}
Sharpe, R.W. , \textit{Differential Geometry}, (Springer, 1997)
\bibitem[RT]{RT}
Reiterer, M. and Trubowitz, E., arxiv.org/abs/0906.3812
\end{thebibliography}
